\documentclass{aa}  

\usepackage{graphicx}
\usepackage{xcolor}
\usepackage{txfonts}
\usepackage[switch]{lineno}

\begin{document} 

 \title{HESS~J1745$-$290 spectrum explained by a transition in the diffusion regime of PeV cosmic rays in the Sgr~A* accretion flow}

   \author{Claudio Muena\inst{1}
          \and
          Mario Riquelme\inst{1}
          \and 
          Andreas Reisenegger\inst{2}
          \and
          Astor Sandoval\inst{1}
          }

   \institute{Departamento de Física, Universidad de Chile, Facultad de Ciencias Físicas y Matemáticas (FCFM), Beauchef 850, Santiago, Chile\\
              \email{cmuena@ing.uchile.cl}
         \and
             Departamento de Física, Facultad de Ciencias Básicas, Universidad Metropolitana de Ciencias de la Educación, Av. José Pedro Alessandri 774, Ñuñoa, Chile\\
             }

   \date{Received XXXX, 2024; accepted XXXX, XXXX}
 
  \abstract 
   {
    The diffuse TeV gamma-ray emission detected in the inner $\sim$ 100 pc of the Galactic Center suggests the existence of a central cosmic-ray accelerator reaching $\sim$ PeV energies. It is interesting to associate this so-called `PeVatron' with the point source HESS~J1745$-$290, whose position is consistent with that of the central supermassive black hole, Sgr~A*. However, the point source shows a spectral break at a few TeV, which is not shown by the diffuse emission, challenging this association.} 
   {
    We seek to build an emission model for the point source consistent with both emissions being produced by the same population of relativistic protons, continuously injected with a power-law spectrum up to $\sim$ PeV energies, near Sgr A*.
    }
   {
   In our model, we assume that the point source is produced by hadronic collisions between the cosmic rays and the gas in the accretion flow of Sgr A*. The cosmic-ray density is calculated taking into consideration cosmic-ray transport due to diffusion and advection, while the properties of the gas are obtained from previous numerical simulations of the accretion flow.}
   {
   Our model succeeds in explaining both the point source and the diffuse emission with the same cosmic rays injected in the vicinity of Sgr A*, as long as the coherence length of the magnetic turbulence in the accretion flow is $l_c\sim(1-3)\times 10^{14}\,\mathrm{cm}$. The spectral break of the point source appears naturally due to an energy-dependent transition in the way the cosmic rays diffuse within the inner $\sim 0.1$ pc of the accretion flow (where most of the emission is produced). 
   }
   {
    Our model supports the idea that Sgr A* can be a PeVatron, whose accelerated cosmic rays give rise to both the point source and the diffuse emission. Future TeV telescopes, like CTAO, will be able to test this model.}

               
  \titlerunning{HESS~J1745$-$290 explained by cosmic rays diffusing out through the accretion flow of Sgr A*}
  
  \authorrunning{Muena et al.}
  
   \maketitle
%

\section{Introduction}
   The very-high energy (VHE) gamma-ray source HESS~J1745$-$290, located at the Galactic Center (GC), has been observed by several imaging atmospheric Cherenkov telescopes (IACTs), such as CANGAROO, VERITAS, HESS and MAGIC (\citealt{Tsuchiya_2004,Kosack_2004,Aharonian_2004,Albert_2006, Abramowski_2016,2018A&A...612A...9H, MAGIC_2020, Adams_2021}). This source is usually characterized by a hard power-law spectrum with a photon index $\sim 2.1$ and an exponential cutoff at $\sim 10$ TeV. However, its spectral shape is also compatible with a broken power-law with the break energy at a few TeV \citep{Aharonian_2009, Adams_2021}.\footnote{We confirm this point in Appendix \ref{ap:fit_hess2016}, using 
   up-to-date HESS data 
   from \cite{Abramowski_2016}.}
   
   Even though several models have been proposed to explain this source, HESS~J1745$-$290 has not yet been associated with any particular astrophysical object. This is mainly because of the limited $\sim 3'$ angular resolution of current IACTs, which translates into $\sim$ 10 pc in projected distance from the GC \citep{2008AIPC.1085..882F}. However, some processes have been proposed: (1) dark matter annihilation in a density cusp at the GC \citep{HooperEtAl2004,Profumo2005,Aharonian_2006DM,Cembranos_2012}, (2) inverse Compton emission by relativistic electrons in the pulsar wind nebula G359.95-0.04 \citep{Hinton_2007,Wang_2006}, and (3) the decay of neutral pions produced by hadronic collisions of cosmic rays (hereafter, CRs) accelerated by the central supermassive black hole Sagittarius A* (Sgr~A*) with background protons in the interstellar gas \citep{Neronov_2005,Chernyakova_2011,RodriguezRamirez_2019}. 

    In this work we focus on the last scenario. Although Sgr~A* is today a relatively quiescent black hole, it has most likely gone through significantly more active phases in the past. This is suggested by the presence of the Fermi bubbles, which are plausibly caused by intense accretion activity some $\sim 10^6-10^7$ years ago \citep{Su_2010}. Also, the observations of X-ray echoes coming from dense gas in the GC region suggest an intense flaring activity in the past few hundred years \citep{Clavel_2013,Marin_2023}. Thus, assuming some efficient accretion-driven CR acceleration mechanism, Sgr~A* may be powering HESS~J1745$-$290 as its accelerated CRs propagate away from the GC and interact hadronically with its surrounding gas.

    The potentially important role of Sgr~A* in producing the TeV emission of HESS~J1745$-$290 is also supported by the diffuse VHE gamma-ray emission from the Central Molecular Zone (CMZ), located in the inner few hundred parsecs of the Galaxy \citep{Abramowski_2016}. This emission is correlated with the density of molecular gas, suggesting a hadronic emission due to CR collisions with background protons. The CR density profile inferred from this assumption is consistent with a $1/r$ dependence (where $r$ is the distance to the GC), supporting a radial diffusion of CRs continuously injected from the inner tens of parsec around the GC. The gamma-ray spectrum of this emission is characterized by a single power-law with index $\sim 2.3$ up to tens of TeV, without a statistically significant spectral cutoff, as shown by H.E.S.S. and VERITAS \citep{Abramowski_2016,Adams_2021}, although some controversy remains due to indications of a spectral turnover around 20 TeV reported by MAGIC \citep{MAGIC_2020}. The possible lack of a cutoff in the diffuse CMZ emission is interesting, since it opens up the possibility for a PeV proton accelerator, or `PeVatron', to be present within the inner tens of parsecs of the Galaxy. These observations have thus been considered as strong indications that Sgr~A* is operating as a PeVatron \citep{Abramowski_2016}, although contributions from other possible CR acceleration systems have also been considered. One possible contribution includes CRs accelerated in the shock front of supernova Sgr A East \citep{Abramowski_2016, Scherer_2022, Scherer_2023}. Another possibility is given by the multiple shocks formed by stellar winds in the Nuclear, Arches and Quintuplet clusters of young massive stars \citep{2019NatAs...3..561A,Scherer_2022,Scherer_2023}, with the Arches and Quintuplet clusters being located at a projected distance of $\sim$ 25 pc from Sgr A*, while the dynamic center of the Nuclear cluster coincides with it \citep{2022ApJ...939...68H}.

Even though the scenario in which Sgr~A* drives the emissions of both HESS~J1745$-$290 and the CMZ is appealing, one of its main challenges is the clear presence of a spectral turnover (either an exponential cutoff or a spectral break) at a few to $\sim 10$ TeV in HESS~J1745$-$290, which appears to be absent from the CMZ emission. One possible way to alleviate this discrepancy is by invoking the absorption of $\gtrsim 10$ TeV gamma rays from HESS~J1745$-$290 by $e^+e^-$ pair production due to their interaction with the interstellar radiation field (ISRF). Although this absorption is not expected to be important given the known ISRF in the GC \citep{Zhang_2005,Aharonian_2009}, its effect may become significant due to clumpiness of the interstellar medium \citep{Guo_2017}. Another possibility is simply a relatively recent decrease in the power of CR injection \citep{Liu_2016}, so the diffuse emission would be powered by CRs up to $\sim $ PeV energies accelerated by Sgr~A* $\gtrsim 10^4$ years ago, while HESS~J1745$-$290 would be associated to a less efficient acceleration period in the last $\sim 10-100$ years. 

In this paper, we present an alternative scenario, in which the spectral turnover of HESS~J1745$-$290 is caused by an energy-dependent transition in the diffusion regime of CRs injected by Sgr~A*. In our model, HESS~J1745$-$290 is caused by CRs experiencing hadronic collisions as they diffuse out through the accretion flow of Sgr~A* ($r\lesssim 0.1$ pc). Even though in our proposed scenario the injected CRs have an energy distribution given by a single power-law up to $\sim$ PeV energies, the transition in their diffusion regime significantly reduces their number density within the accretion flow for energies $\gtrsim 20$ TeV. This density reduction in turn imprints a break in the gamma-ray emission spectrum at a few TeV. 
 
Our proposed model needs three basic ingredients: 
\begin{itemize}
    \item The radial profiles for the density, velocity and magnetic field strength in the gas at $r\lesssim 1$ pc, which has been proposed to be mainly sourced by the wind of $\sim$ 30 Wolf-Rayet (WR) stars, which belong to the Nuclear cluster and orbit around Sgr~A*, at $\sim 0.1 - 1$ pc from it  \citep{Cuadra_2006,Cuadra_2007}. These profiles are obtained from previously reported  hydrodynamic \citep{Ressler_2018} and magneto-hydrodynamic \citep[MHD;][]{Ressler_2020_small_B} simulations of the Sgr A* accretion flow. Interestingly, despite the relatively large magnetic fields used for the stellar winds in the MHD studies, their obtained angle-averaged gas density, temperature and velocity are largely insensitive to the assumed magnetic field strength and are essentially the same as the ones found in the hydrodynamic simulations \citep{Ressler_2020_small_B}. Because of this, below we refer to both hydrodynamics and MHD results when describing the assumed gas behavior in our model.
    
    \item A second ingredient is given by the diffusion and advection properties of the CRs within the MHD turbulence in the $r\lesssim 1$ pc region. While the properties of the gas flow are obtained from previous hydrodynamic and MHD simulations, the CR diffusion properties are inferred from previous studies of test particle CRs propagating in synthetic turbulence, which we assume to be strong (i.~e., the mean, large-scale magnetic field is much weaker than the fluctuating field), isotropic, and with a Kolmogorov spectrum \citep{Subedi_2017,Mertsch_2020,Dundovic_2020}. Assuming a mean field much weaker than the turbulent field is one of the main assumptions in our work, which allows us to neglect a possible anisotropy in the CR diffusion, as it has also been found in test particle simulations \citep[see, e.g.,][]{2022MNRAS.514.2658R,2022SNAS....4...15R,2023arXiv231101497R}.
    
    \item Finally, we assume a flux of centrally injected CRs whose spectral distribution corresponds to a single power-law whose index, normalization, and maximum energy (reaching $\sim$PeV energies) are consistent with estimates obtained from the diffusive CMZ emission \citep{Abramowski_2016}. 
\end{itemize}  

Most parameters in our model are thus relatively well constrained by previously reported hydrodynamic and MHD simulations of the Sgr~A* accretion flow, as well as by the spectral properties of the CMZ emission. The only exception is the coherence length $l_c$ of the MHD turbulence in the accretion flow, which, despite playing a critical role in the diffusion properties of the CRs, is a poorly known parameter of our model. Interestingly, we find that by choosing $l_c \sim (1-3)\times 10^{14}$ cm, our model is capable of matching fairly well the main spectral features of HESS~J1745$-$290, including its spectral turnover, while being consistent with the injected CR spectrum required to reproduce the CMZ emission.

We note that, in our model, the CR injection occurs within a radius $r\lesssim 10 R_g$, where $R_g$ is the gravitational radius of the black hole (or Schwarzschild radius, which in the case of Sgr A* is $R_g\approx 10^{12}$ cm). This is justified by the fact that electron acceleration events in Sgr A* are routinely inferred in association with near-infrared and X-ray flares \citep[likely powered by reconnection events;][]{2017MNRAS.468.2447P,2020ApJ...898..138S,2022MNRAS.511.3536S}, whose emission regions have been located at a few $R_g$ from the central black hole \citep{2018A&A...618L..10G}. Although this injection region may itself be the source of significant gamma-ray emission \citep{RodriguezRamirez_2019}, our work focuses solely on the emission produced at $r\gtrsim 10 R_g$ (which, as we show, is in fact dominated by $r\gg 10R_g$). 

Our paper is organized as follows. In \S~\ref{sec:crdensity}, we calculate the CR density as a function of radius and energy within the accretion flow of Sgr~A*, considering the diffusion properties of CRs in that region, as well as the velocity profile of the background gas. In \S~\ref{sec:gamma}, we calculate the gamma-ray emission resulting from the hadronic interaction of the CRs with the background gas. In \S~\ref{sec:consistency}, we verify the consistency of the assumptions made in our calculations. In \S~\ref{sec:Discussion}, we provide a discussion regarding CR diffusion time-scales and how they could be used to observationally test our model. Finally, our conclusions are presented in \S~\ref{sec:Conclusions}.


\section{Cosmic ray density model}
\label{sec:crdensity}

In this work we assume that CRs are ultra-relativistic protons being steadily injected near the event horizon of Sgr~A* and propagate to larger radii $r$ through its accretion flow. This accretion flow originates mainly from the gas being injected by a cluster of WR stars orbiting at $\sim 0.1 - 1$ pc from Sgr~A* \citep{Cuadra_2006,Cuadra_2007}.
The CR transport equation can be written as \citep[e.g.,][]{2007ARNPS..57..285S}

\begin{linenomath}
\begin{equation}
\label{eq:diffadv}
\begin{aligned}
    \frac{\partial}{\partial t}\Bigg(\frac{dn_{CR}}{dE}(E,\mathbf{r},t)\Bigg)=& S(E,\mathbf{r},t) + \nabla \cdot \Bigg\{D(E,\mathbf{r})\nabla\Bigg(\frac{dn_{CR}(E,\mathbf{r},t)}{dE}\Bigg)  \\
    &  -\mathbf{v}(\mathbf{r},t)\frac{dn_{CR}(E,\mathbf{r},t)}{dE}\Bigg\},
\end{aligned}
\end{equation}
\end{linenomath}
where $\mathrm{d}n_{\mathrm{CR}}/\mathrm{d}E(E,\mathbf{r},t)$ is the CR density per unit energy $E$ as a function of position $\mathbf{r}$ and time $t$. The term $S(E,\mathbf{r},t)$ is the source term that quantifies CR injection, $D(E,\mathbf{r})$ is the CR diffusion coefficient and $\mathbf{v}(\mathbf{r},t)$ is the gas velocity, which allows advection to contribute to CR propagation. In Eq. \ref{eq:diffadv} we have neglected possible CR energy gains (e.g., due to stochastic acceleration) and energy losses (we verified that energy losses are indeed small by comparing the CR diffusion times found in \S \ref{sec:Discussion} with the cooling times due to hadronic CR interactions). 
For simplicity, and as a first approach to this problem, in this work we solve Eq. \ref{eq:diffadv} assuming a stationary and spherically symmetric CR density model. In what follows, we describe the parameters and assumptions regarding CR injection and transport. 

\subsection{Cosmic ray injection}
\label{sec:crinject}

In our model, we assume that the CR injection occurs only within a radius $r < 10 R_g$, and solve for $\mathrm{d}n_{\mathrm{CR}}/\mathrm{d}E(E,r)$ at $r > 10 R_g$. Thus, our calculations do not require the exact form of the source term $S(E,r)$ in Eq. \ref{eq:diffadv}, but only the total number of injected CRs per unit energy $E$ and per unit time $t$ within the injection region, which is given by
\begin{linenomath}
\begin{equation}
\label{eq:normal}
\mathrm{d}N/\mathrm{d}E\mathrm{d}t(E)=4\pi\int_0^{10R_g}dr\,r^2 S(E,r).   
\end{equation}
\end{linenomath}
For this, we assume a power-law dependence,
\begin{linenomath}
\begin{equation}
\label{eqn:source}
\frac{\mathrm{d}N}{\mathrm{d}E\mathrm{d}t}(E) = 2.3\times 10^{36} \, f(q,E_{\textrm{max}}) \:\hat{Q} \, \left(\frac{E}{1\: \mathrm{TeV}}\right)^{-q} \, \mathrm{erg}^{-1}\mathrm{s}^{-1} \mathrm{,}
\end{equation}
\end{linenomath}
for $E$ smaller than a maximum energy $E_{\textrm{max}}$, and $\mathrm{d}N/(\mathrm{d}E\mathrm{d}t)=0$ otherwise. Here,
$q$ is the spectral index of the injected CR spectrum, $\hat{Q}$ is a parameter that quantifies the injection power, and
\begin{linenomath}
\begin{equation}
\label{Emax}
f(q,E_{\textrm{max}}) = \frac{2-q}{\left(\frac{E_{\textrm{max}}}{1\: \mathrm{TeV}}\right)^{2-q}-10^{2-q}} \,\,\,\,\,\,\,\, \textrm{if }q\neq 2, 
\end{equation}
\end{linenomath}
while, if $q=2$, 
\begin{linenomath}
\begin{equation}
    f(q=2,E_{\textrm{max}})=\left[\ln\left(\frac{E_{\textrm{max}}}{1 \,\textrm{TeV}}\right)-\ln(10)\right]^{-1},
\end{equation}
\end{linenomath}
so the CR injection power for $E\geq 10~\mathrm{TeV}$ is
\begin{linenomath}
\begin{equation}
Q(E>10\,\textrm{TeV}) = 6\times 10^{36} \,\hat{Q}\, \textrm{erg s}^{-1}.
\label{eq:Q}
\end{equation} 
\end{linenomath}

\subsection{Cosmic ray transport}
\label{sec:crtransport}

Since our model is stationary and assumes spherical symmetry, integrating Eq. \ref{eq:diffadv} over a spherical volume of radius $r$, and considering the definition of $\mathrm{d}N/\mathrm{d}E\mathrm{d}t(E)$ provided by Eq. \ref{eq:normal}, we obtain
\begin{linenomath}
\begin{equation}
\label{eqn1: diff}
\frac{\mathrm{d}N}{\mathrm{d}E\mathrm{d}t}(E)=4\pi r^2\Bigg(v_{\textrm{gas}}(r)\: \frac{\mathrm{d}n_{\textrm{CR}}}{\mathrm{d}E}(E,r)-D(E,r) \frac{\partial}{\partial r} \left(\frac{\mathrm{d}n_{\mathrm{CR}}}{\mathrm{d}E}(E,r)\right)\Bigg),
\end{equation}
\end{linenomath}
where $v_{\textrm{gas}}(r)$ is the angle-averaged gas velocity in the radial direction at a radius $r$. We see that both diffusion and advection are expected to contribute to CR propagation. However, below we show that, in our model, diffusion is the dominant process at $r\lesssim 0.07$ pc for all energies of interest, while for $r\gtrsim 0.07~\mathrm{pc}$ advection dominates for CRs with low enough energies. Thus, as an approximation, in our calculations of $\mathrm{d}n_{\mathrm{CR}}/\mathrm{d}E(E,r)$ from Eq.~(\ref{eqn1: diff}) we impose that, at a given radius $r$ and energy $E$, CRs are transported entirely by one process, either diffusion or advection (i.e., keeping only one of the two terms on the right hand side of Eq.~(\ref{eqn1: diff})), depending on which one is more efficient. This is done by comparing the radial gas velocity $v_{\mathrm{gas}}$ to the diffusion velocity
\begin{linenomath}
\begin{equation} \label{eq. vdiff}
v_{\text{diff}}(E,r)\equiv\frac{D(r,E)}{r},
\end{equation}
\end{linenomath}
and neglecting the diffusion (advection) term on the right hand side of Eq.~(\ref{eqn1: diff}) if $v_{\mathrm{gas}}$ ($v_{\text{diff}}$) is larger.\footnote{Comparing $v_{\mathrm{gas}}$ and $v_{\text{diff}}$ is equivalent to comparing the corresponding advective and diffusive escape times of CRs from a region of size $r$, which are respectively given by $\tau_{\textrm{adv}} = r/v_{\text{gas}}$ and  $\tau_{\textrm{diff}} = r/v_{\text{diff}}$.}
We note that CRs can also be advected by the turbulent motions of the gas, giving rise to turbulent diffusion. We show below that this effect can be neglected for the cases of interest (see \S~\ref{sec:consistency}), so in our model CR transport is determined by the properties of $D(E,r)$ and $v_{\textrm{gas}}(r)$. 

\subsubsection{Diffusion model}
\label{sec:diffmodel}

In order to model $D(E,r)$, we use results from previous test particle simulations of CRs propagating in synthetic MHD turbulence (i.e., magnetostatic turbulence with a pre-specified spectrum), which we assume to be strong in the sense that the rms magnetic field magnitude is much larger than any large-scale mean field, isotropic and with a Kolmogorov spectrum \citep{Subedi_2017,Mertsch_2020,Dundovic_2020}. This turbulence is characterized by a coherence length $l_c$, which is the spatial scale in which most of the turbulent magnetic energy is contained \citep{fleishman}. These simulations have identified two main diffusion regimes, which are also consistent with theoretical diffusion models \citep{Aloisio_2004,Subedi_2017}. The ``high-energy diffusion'' (HED) regime occurs when the energy of the CRs is such that their Larmor radii $R_L$ satisfy $R_L \gg l_c$ ($R_L=E/e B$, where $e$ is the proton electric charge and $B$ is the magnitude of the magnetic field), while the ``low-energy diffusion'' (LED) regime occurs when $R_L\ll l_c$. We obtain expressions for the effective mean free path $\lambda_{\textrm{mfp}}$ of CRs in these two diffusion regimes from \cite{Subedi_2017} (where a broken power law shape is assumed for the turbulence spectrum). Figure 1 of \cite{Subedi_2017} shows that, for $R_L \ll l_c$ (LED), $\lambda_{\textrm{mfp}}$ is given by
\begin{linenomath}
\begin{equation}
\lambda_{\textrm{mfp}} \approx 0.4\,l_c^{2/3}R_L^{1/3},
\label{eq:lemfp}
\end{equation}
\end{linenomath}
while, for $R_L \gg l_c$ (HED),
\begin{linenomath}
\begin{equation}
\lambda_{\textrm{mfp}} \approx 2\,l_c^{-1}R_L^{2},
\label{eq:hemfp}
\end{equation}
\end{linenomath}
with a smooth transition for $R_L \sim l_c$. (In Appendix~\ref{ap:heuristic} we give heuristic derivations for the dependence on $l_c$ and $R_L$ in these two limiting regimes.) Notice that, since Eqs. \ref{eq:lemfp} and \ref{eq:hemfp} are approximate expressions obtained from Fig. 1 of \cite{Subedi_2017}, their normalizations should only be considered to have an accuracy at the $\sim 10\%$ level. Another approximation is that, in our calculations, the transition in the $\lambda_{\textrm{mfp}}$ behavior at $R_L \sim l_c$, is satisfied by assuming 
\begin{linenomath}
\begin{equation}   
\lambda_{\textrm{mfp}} =
        \begin{cases}
            0.4 \: l_c^{2/3} R_L^{1/3} &\quad\text{for}\quad R_L \leq 0.38 \: l_c, 
            \\ 
            2 \: l_c^{-1} R_L^{2} &\quad\text{for}\quad R_L > 0.38 \: l_c.\\
        \end{cases}
        \label{eq:lambdamfp}
\end{equation}
\end{linenomath}
Thus, in order to obtain $D(E,r)=\lambda_{\textrm{mfp}}(E,r)c/3$ (where $c$ is the speed of light), 
we need an expression for the rms magnitude of the magnetic field $B$ as a function of $r$. Based on the MHD simulations of the Sgr~A* accretion flow of \citet{Ressler_2020_small_B}, we assume it to have a power-law profile,
\begin{linenomath}
\begin{equation}
\label{eqn: B}
B(r) = 10 \: \hat{B}\: \left(\frac{r}{10 R_g}\right)^{-n} \textrm{G},
\end{equation}
\end{linenomath}
where $n\approx 1$, while $\hat{B}$ is a factor of order unity, fairly independent of the conditions assumed for the wind of the WR stars. From this, we obtain a space and energy dependent Larmor radius
\begin{linenomath}
\begin{equation}
\label{eqn: R_L}
R_L \approx 3.3\times 10^8 \: \hat{B}^{-1}\: \frac{E}{\mathrm{TeV}}\: \left(\frac{r}{10 R_g}\right)^n \textrm{cm}.
\end{equation}
\end{linenomath}
In contrast, the value of the coherence length $l_c$ in the accretion flow is very uncertain. In order to model the possible radial dependence of this parameter, we also assume a power-law shape for it, with 
\begin{linenomath}
\begin{equation}
\label{eqn: lc}
l_c(r) = 10^{14} \: \hat{l}_c \left(\frac{r}{0.07 \,\textrm{pc}}\right)^{m} \textrm{cm},
\end{equation} 
\end{linenomath}
where $\hat{l}_c$ and $m$ are free parameters of our model. 

Thus, the diffusion coefficient in the LED regime, $R_L \ll l_c$, is
\begin{linenomath}
\begin{equation}
 \begin{aligned}
 D_{\mathrm{LED}}(E,r) = & \, 6 \times 10^{21} \, (2.1\times 10^4)^{-2m/3} \, \hat{l}_c^{2/3} \,\hat{B}^{-1/3} \\
 & \times\left(\frac{E}{1\text{TeV}}\right)^{1/3} \, \left(\frac{r}{10R_g}\right)^{\frac{1}{3}(n+2m)} \, \text{cm}^{2} \text{s}^{-1},\\
 \end{aligned}
\label{eq:D1}
\end{equation}
\end{linenomath}
while in the HED regime, $R_L \gg l_c$,
\begin{linenomath}
\begin{equation}
 \begin{aligned}
 D_{\mathrm{HED}}(E,r) = & \, 2 \times 10^{13} \, (2.1\times 10^4)^{m} \, \hat{l}_c^{-1} \, \hat{B}^{-2} \\
 & \times\left(\frac{E}{1\text{TeV}}\right)^2 \, \left(\frac{r}{10R_g}\right)^{2n-m} \, \text{cm}^{2} \text{s}^{-1}.\\
 \end{aligned}
\label{eq:D2}
\end{equation}
\end{linenomath}
\subsubsection{Advection model}
\label{Gass prop}
In order to model $v_{\textrm{gas}}(r)$, we need to characterize the gas dynamics within the Sgr~A* accretion flow. For this, we use the hydrodynamic simulations presented by \cite{Ressler_2018}, which assume that the gas is mainly provided by the wind of a cluster of WR stars orbiting at $0.1-1$ pc from Sgr~A*. Interestingly, these simulations show no significant differences with the MHD simulations of \citep{Ressler_2020_small_B}, when angle-averaged quantities are considered. For simplicity, we assume that this gas is a plasma composed of electrons and protons of equal number densities $n_e(r)=n_p(r)=n_{\textrm{gas}}(r)$. The profile of the gas density $n_{\textrm{gas}}$ is obtained from Fig. 11 of \cite{Ressler_2018}, which provides quantities averaged over 
angle and time (over the 100 years previous to the present day). This density profile can be approximated by\footnote{The first of these relations is nearly identical to that inferred by \citet{Eatough_2013} from the X-ray observations reported by \citet{Muno_2004}.}
\begin{linenomath}
\begin{equation} 
    \label{eq. gas dens}
    n_{\textrm{gas}}(r)=\left\{\begin{matrix}
 25 \left(\frac{r}{0.4\text{pc}}\right)^{-1} \text{cm}^{-3}& \:\:\: \mathrm{if}\:\:  r\leq \:0.4\: \mathrm{pc,}\\ 
25 \left(\frac{r}{0.4\text{pc}}\right)^{-2} \text{cm}^{-3}& \:\:\: \mathrm{if} \: \:r > \:0.4 \: \mathrm{pc.}
\end{matrix}\right.
\end{equation}
\end{linenomath}
Notice that, according to Fig. 11 of \citealt{Ressler_2018}, the $n_{\textrm{gas}}(r)$ profile at $0.07\, \textrm{pc}\lesssim r \lesssim 0.4\, \textrm{pc}$ is steeper than $n_{\textrm{gas}}(r) \propto r^{-1}$. However, we show in \S \ref{sec:qualitative} that the gamma-ray emission from this region is expected to be 
$\lesssim 20-30\%$ of the total emission at all energies. Therefore, as a reasonable approximation, we use Eq. \ref{eq. gas dens} to describe $n_{\textrm{gas}}(r)$ at all radii smaller than $0.4$ pc.

Regarding the gas velocity, \cite{Ressler_2018} identifies four regions within the accretion flow:

\begin{itemize}
    \item Region I: This is the outflow-dominated region, defined by $r > 0.4$ pc. 
    Here, the gas moves away from the GC and $v_{\textrm{gas}}(r)$ is essentially given by the speed of the WR stars' winds.
    \item Region II: Located at $0.07 \: \mathrm{pc} < r < 0.4$ pc, this region corresponds to the `feeding' region, where most of the gas injection from the WR stars' winds occurs, and in which the gas motion predominantly points away from the GC.
    \item Region III: Corresponds to a `stagnation' region at $0.01 \mathrm{pc} < r < 0.07$ pc, which is characterized by 
    inflows and outflows of gas, with the net mass accretion rate nearly vanishing.
    \item Region IV: Corresponds to an inflow-dominated region at $r < 0.01$ pc, where the mass accretion rate $\dot{M}$ is roughly constant.     
\end{itemize}

In region IV, we consider a constant mass accretion rate $\dot M\approx 10^{-8} M_\odot/\mathrm{yr}$ \citep{Ressler_2020,Dexter_2020}, with which we find that 
\begin{linenomath}
\begin{align}
\label{eqn:vr}
    v_{\text{gas}}(r) &= -\frac{\dot{M}}{4\pi r^2 n_{\textrm{gas}}(r) m_p} \nonumber\\
    &\approx -3\times 10^4 \left(\frac{r}{0.01\mathrm{pc}}\right)^{-1} \: \mathrm{cm/s},
\end{align}
\end{linenomath}
where $m_p$ is the proton mass.

In region III, the average radial velocity of the gas tends to cancel out, therefore CR transport should not be significantly affected by advection. In what follows, we neglect the effect of $v_{\text{gas}}(r)$ for regions III and IV, for which we assume that CR transport is dominated by diffusion. The validity of this assumption is verified in \S~\ref{sec:consistency}.  

In regions I and II ($r > 0.07$ pc), we can estimate $v_{\text{gas}}$ from Fig. 18 in \cite{Ressler_2018}, which can be approximated by
\begin{linenomath}
\begin{equation}
    \label{eq. v_ad}
    v_{\textrm{gas}}(r)\approx
    \begin{cases}
        700(r/0.4 \mathrm{pc}) \: \mathrm{km/s} \:\:\:\: \mathrm{for }\: 0.07 \: \mathrm{pc }<r<0.4 \:\mathrm{pc}\\
        700 \: \mathrm{km/s} \:\:\:\:\:\:\:\:\:\: \:\:\:\:\:\:\:\:\:\:\:\:\:  \mathrm{for }\: \: r>0.4 \:\mathrm{pc.}\\
    \end{cases}
\end{equation}
\end{linenomath}
In this case, $v_{\textrm{gas}}$ dominates CR transport for low enough energies, while diffusion dominates for higher energy particles. Thus, for $0.07\,\mathrm{pc}< r < 0.4\,\mathrm{pc}$, we consider diffusive transport at energies for which $v_{\mathrm{diff}}>v_{\mathrm{gas}}$, and advective transport in the opposite case. For simplicity, at $r>0.4\,\mathrm{pc}$, we assume the same transport mechanism that dominates at $r=0.4\,\mathrm{pc}$, which, depending on $E$, could either be HED 
or advection. As discussed in \S~\ref{sec:transport>0.4}, this should not affect our main results.

\subsection{Previous constraints on parameters}
\label{sec:approxeffect}

Since our purpose is to build a model for the point source HESS~J1745$-$290 compatible with the diffuse emission from the CMZ, the injection parameters $q$, $\hat{Q}$ and $E_{\textrm{max}}$ have to obey restrictions from the CMZ emission. This emission has been found to be consistent with a CR proton spectral index of $\Gamma_p \approx 2.3$ \citep{Abramowski_2016}. This, however, does not correspond to the index $q$ of the injected CRs, since the CR spectrum in the CMZ is also affected by the energy-dependent residence time of the CRs in this region, which is inversely proportional to the CR diffusion coefficient in the CMZ, assumed to have a power-law shape, $D_{\textrm{CMZ}}(E)\propto E^{\delta}$. This implies that $\Gamma_p = q + \delta$, with the value of $\delta$ likely between $\delta \approx 1/3$ and 1/2, as expected for Kolmogorov and Kraichnan types of magnetic turbulence, respectively \citep{Aharonian_2004_VHE_book}. Thus, a reasonable expectation for $q$ is $ \approx 2$. Also, making different assumptions regarding the magnitude of $D_{\textrm{CMZ}}(E)$, \cite{Abramowski_2016} and \cite{Scherer_2022} found the CR injection power for $E>10$ TeV to be $Q(E>10\,\mathrm{TeV})\approx 8\times 10^{37} \mathrm{erg}\: \mathrm{s}^{-1}$ and 
$3\times 10^{36} \mathrm{erg}\: \mathrm{s}^{-1}$, implying $\hat{Q} \approx 13$ and 0.5, respectively. Additionally, the spectrum of the diffuse emission restricts the maximum CR energy to $E_{\textrm{max}} \gtrsim 1$ PeV \citep{Abramowski_2016, Adams_2021}.

Further parameter restrictions can be obtained from previous MHD simulations of the Sgr~A* accretion flow. For instance, \cite{Ressler_2018} shows that assuming a beta parameter in the winds of the WR stars $\beta_W=10^4$, the magnetic field near Sgr~A* reaches the value $B(r=10R_g)\approx 10$ G ($\hat{B}\approx 1$), and the parameter $n$ (defined in Eq.~(\ref{eqn: B})) is approximately $n \approx 1$. In the case of $\beta_W=10^2$ they obtain essentially the same $\hat{B}\approx 1$, while $B(r=0.1\,\mathrm{pc})$ is a factor $\sim 4$ larger than in the $\beta_W=10^4$ case. For our assumed power-law dependence $B\propto r^{-n}$ (Eq.~(\ref{eqn: B})), the latter implies $n\approx 0.9$.

Given these considerations, we will restrict our analysis to cases with $\hat{B} = 1$ and $q=2$, and will consider two possible values for $n$, namely $n=1$ and 0.9. We will show that, with these choices, our model can reproduce the main features of the emission from HESS~J1745$-$290 quite well, if $\hat{Q} \approx 6-13$ and $E_{\textrm{max}} \gtrsim 1$ PeV, which respects the restrictions from the diffuse CMZ emission provided by \cite{Abramowski_2016}. Regarding the more uncertain parameters $\hat{l}_c$ and $m$, we will show that, fitting the data of HESS~J1745$-$290 is strongly dependent on $\hat{l}_c$ being in the range $\hat{l}_c \approx 1-3$ (for $n=1$ and 0.9, respectively), while $m$ must be small ($m \lesssim 0.3$). In other words, the coherence length $l_c$ must depend only weakly on $r$.

\begin{figure}[h]
    \centering
    \includegraphics[width=0.48 \textwidth]{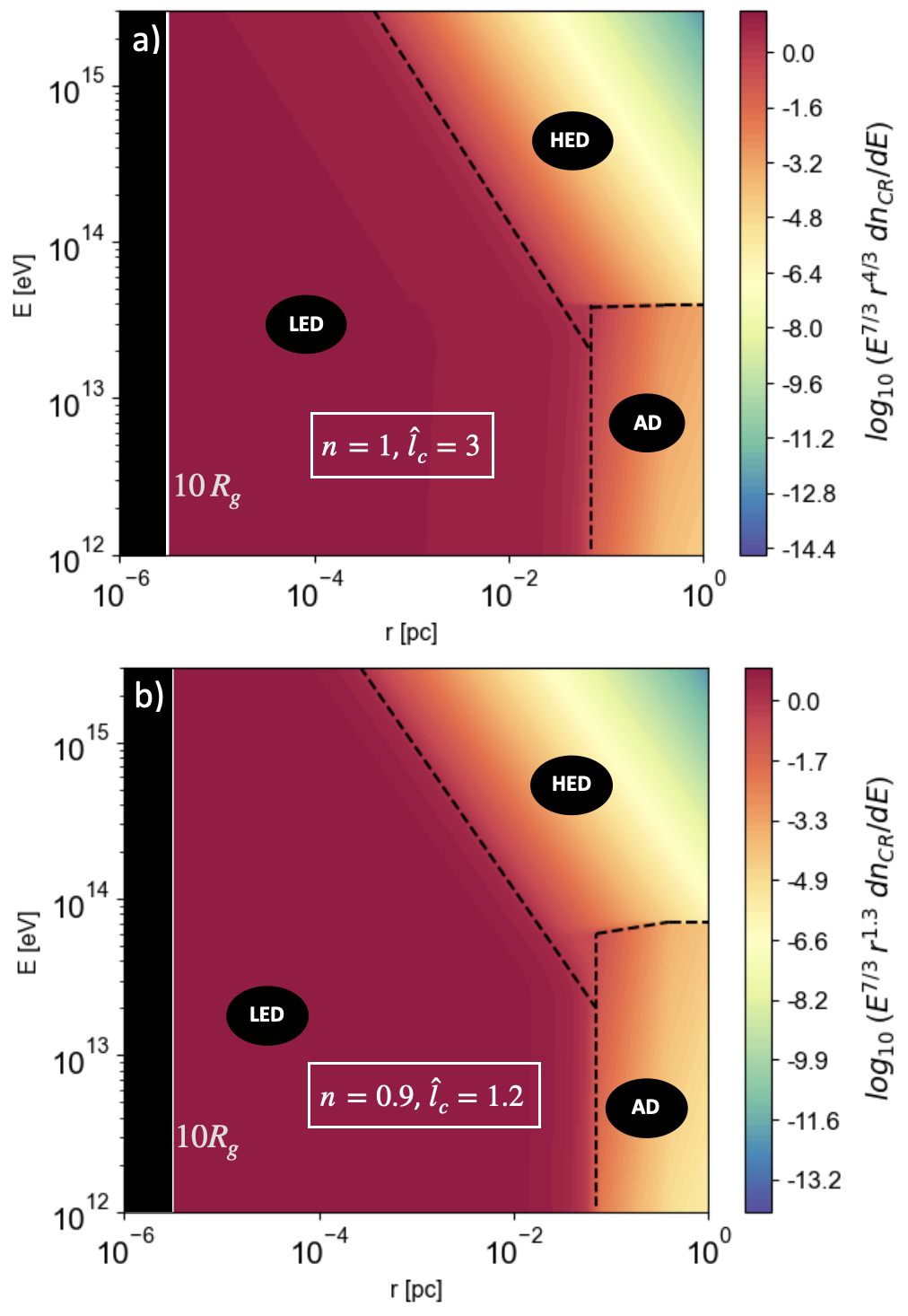}
      \caption{ \small  Cosmic ray density per unit energy $dn_{\textrm{CR}}/dE$ as a function of the radial coordinate $r$ (horizontal axis) and the CR energy $E$ (vertical axis) for cases \textit{2} (panel $a$) and \textit{4b} (panel $b$) specified in Table \ref{tab:param}. The color scale gives $\log_{10}(E^{7/3}r^\xi\mathrm{d}n_{\mathrm{CR}}/\mathrm{d}E)$ in cgs units (with $\xi=4/3$ in panel \textit{a} and $\xi=1.3$ in panel \textit{b}), and the dashed lines mark the boundaries between the three CR transport regimes discussed in the text, namely the low-energy diffusion (LED), high-energy diffusion (HED), and advection (AD) regimes.}
    \label{fig:CR_dens}
\end{figure}

\subsection{Cosmic ray density profiles}

Considering our assumptions regarding CR injection and transport described in \S~\ref{sec:crinject} and \S\ref{sec:crtransport}, we solve Eq.~(\ref{eqn1: diff}) assuming continuity in $\mathrm{d}n_{\mathrm{CR}}/\mathrm{d}E$ at the transition between regions dominated by diffusion and by advection.\footnote{The existence of discontinuities in $\mathrm{d}n_{\mathrm{CR}}/\mathrm{d}E$ as a function of $r$ would be inconsistent with the presence of diffusion, which is present in all regions, even in those where advection dominates.} Figure \ref{fig:CR_dens} shows examples of the CR density profile for two combinations of parameters (Cases \textit{2} and \textit{4b}  in Table~\ref{tab:param}), where we can identify the three main transport regimes mentioned above:
\begin{itemize}
    \item Low-energy diffusive (LED) regime: In this region, CR transport is dominated by diffusion, with $\lambda_{\textrm{mfp}}$ in the $R_L < 0.38\,l_c$ regime of Eq.~(\ref{eq:lambdamfp}). Solving Eq.~(\ref{eqn1: diff}) in that regime, we obtain:
\begin{linenomath}
\begin{equation}
\label{Eq: CR dens1}
    \begin{aligned}
     \frac{\mathrm{d}n_{\mathrm{CR}}}{\mathrm{d}E}(E,r)= & \,3 \: (2.1\times 10^4)^{2m/3}\,\hat{Q} \,  \hat{l}_c^{-2/3}\,\hat{B}^{\frac{1}{3}} \,\frac{f(q,E_{\textrm{max}})}{1+\frac{1}{3}(n+2m)}\\
     &   \times \, \left(\frac{E}{1 \text{TeV}}\right)^{-q-\frac{1}{3}} \left(\frac{r}{10 R_g}\right)^{-(1+\frac{1}{3}(n+2m))} \:\text{erg}^{-1}\text{cm}^{-3} . 
     \end{aligned}  
\end{equation}
\end{linenomath}
For simplicity, Eq.~(\ref{Eq: CR dens1}) does not include the terms that ensure continuity of $\mathrm{d}n_{\mathrm{CR}}/\mathrm{d}E$ at the boundaries between regions dominated by different transport regimes, which are subdominant. These continuity terms are, however, included in the calculation shown in Fig. \ref{fig:CR_dens} as well as in the gamma-ray emission calculations shown in \S~\ref{sec:gamma}. In Appendix \ref{ap:fullexpressions}, we provide the full expression for $\mathrm{d}n_{\mathrm{CR}}/\mathrm{d}E$ in the LED regime, including the boundary terms. 

    \item High-energy diffusive (HED) regime: In this case, CR transport is also dominated by diffusion, but with $\lambda_{\textrm{mfp}}$ in the $R_L > 0.38\,l_c$ regime of Eq.~(\ref{eq:lambdamfp}). In this case, the CR density is given by 
\begin{linenomath}
\begin{equation}
\label{Eq: CR dens2}
    \begin{aligned}
     \frac{\mathrm{d}n_{\mathrm{CR}}}{\mathrm{d}E}(E,r)= & \, 10^{9}  \,(2.1\times 10^4)^{-m}\: \hat{Q} \: \hat{l}_c \:  \hat{B}^{2}\, \frac{f(q,E_{\textrm{max}})}{1+2n-m} \\
     & \times \, \left(\frac{E}{1 \text{TeV}}\right)^{-q-2}\left(\frac{r}{10 R_g}\right)^{-(1+2n-m)}\: \text{erg}^{-1}\text{cm}^{-3},\\
     \end{aligned}  
\end{equation}
\end{linenomath}
dropping off much faster with increasing $r$ (or increasing $E$), as seen in the upper right corner of Fig.~\ref{fig:CR_dens}. (Here, we are also excluding subdominant terms that ensure continuity of $\mathrm{d}n_{\mathrm{CR}}/\mathrm{d}E$, but we provide the full expression in Appendix \ref{ap:fullexpressions}.)

    \item Advection-dominated (AD) regime: In this regime, CR propagation is dominated by advection, and their density is given by 
\begin{linenomath}
\begin{equation}
    \label{eqn:Adv1}
    \begin{aligned}
\frac{\mathrm{d}n_{\mathrm{CR}}}{\mathrm{d}E}(E,r)= & \,3\times 10^6 \: \hat{Q} \, f(q,E_{\textrm{max}})\,\left(\frac{E}{1\text{TeV}}\right)^{-q}\\
& \times \, \left(\frac{r}{10R_g}\right)^{-3} \: \text{erg}^{-1}\text{cm}^{-3}
    \end{aligned}
\end{equation}
\end{linenomath}
for $0.07$ pc $\leq r \leq$ $0.4$ pc, and
\begin{linenomath}
\begin{equation}
    \label{eqn:Adv2}
    \begin{aligned}
\frac{\mathrm{d}n_{\mathrm{CR}}}{\mathrm{d}E}(E,r)= &\,21 \:  \hat{Q} \, f(q,E_{\textrm{max}})\,\left(\frac{E}{1\text{TeV}}\right)^{-q}\\
& \times \, \left(\frac{r}{10R_g}\right)^{-2}\, \text{erg}^{-1}\text{cm}^{-3} 
    \end{aligned}
\end{equation}
\end{linenomath}
for $r > 0.4$ pc, where again in both cases the CR density drops off much faster with increasing $r$ than in the LED regime, as seen in the lower right corner of Fig. \ref{fig:CR_dens}. (No boundary terms were neglected in Eqs.~(\ref{eqn:Adv1}) and (\ref{eqn:Adv2}).)
\end{itemize}
As seen in Figure~\ref{fig:CR_dens}, high-energy CRs diffusing out from the central source undergo a transition from the 
LED regime to the 
HED regime at an energy-dependent critical radius $r_C$ determined by the condition $R_L= 0.38\, l_c$, yielding
\begin{linenomath}
\begin{equation}
    \label{eqn:crit radius}
    r_C(E)=0.07 \left[\frac{1.3\times 10^5 \: \hat{l}_c \: \hat{B}}{(2.1\times 10^4)^n}\right]^{\frac{1}{n-m}} \left(\frac{E}{\text{TeV}}\right)^{-\frac{1}{n-m}} \, \text{pc},
\end{equation}
\end{linenomath}
which is marked by the diagonal dashed lines in the high energy part of Figs.~\ref{fig:CR_dens}$a$ and \ref{fig:CR_dens}$b$. At lower energies, instead, CRs transit from the LED regime to the advective regime at a fixed radius $r=0.07\,\mathrm{pc}$. The energy separating these two types of transitions, $E_C$, is determined by $r_C(E_C)=0.07\,\mathrm{pc}$ (corresponding to the inner boundary of the feeding region), which yields 
\begin{linenomath}
\begin{equation}\label{eqn: E_crit}
    E_C=6.2\,(2.1\times 10^4)^{1-n}\,\hat l_c\,\hat B \,\,\mathrm{TeV.}
\end{equation}
\end{linenomath}
Figures ~\ref{fig:CR_dens}$a$ and \ref{fig:CR_dens}$b$ also show that, at energies $E > E_C$, advection continues to be dominant over HED 
for $r>0.07$ pc until an energy $E^* \sim (2-3) E_c$, after which HED 
dominates at $r > 0.07$ pc. Notice that the maximum energy at which advection dominates for $r>0.07$ pc is constant in Case $2$ (Fig. ~\ref{fig:CR_dens}$a$), but it has a weak dependence on $r$ in Case $4b$ (Fig. ~\ref{fig:CR_dens}$b$). This is because, in the range $0.07 \textrm{ pc}<r<0.4 \textrm{ pc}$, making $v_{\textrm{gas}}(r)$ equal to $v_{\textrm{diff}}(E,r)$ in the HED regime, we obtain
\begin{linenomath}
\begin{equation}
    \label{eq.v_gas=v_diff}
    \frac{E}{\textrm{TeV}}= 15.8\,(2.1\times 10^4)^{-m/2}\hat{l}_c^{1/2}\hat{B}\,\Bigg(\frac{r}{10R_g}\Bigg)^{1-n+m/2}.
\end{equation}
\end{linenomath}
This implies that in Case 2 ($n=1$, $m=0$), the limit between the AD and HED regimes is simply given by $E=E^*$, while in Case $4b$ ($n=0.9$, $m=0$), this energy becomes $E \propto r^{0.1}$. For $r > 0.4$ pc, we have assumed, for simplicity, that CR transport is dominated by the same process that dominates at $r=0.4$ pc (which explains why in Figure \ref{fig:CR_dens}$b$ the energy that divides the HED and AD regimes becomes constant at $r > 0.4$ pc). We show in \S \ref{sec:transport>0.4} that this approximation does not affect our results, given that emission at $r > 0.4$ pc is negligible in our model. 

\begin{table}[h!]
\centering
 \small
\begin{tabular}{||c |c| c| c| c| c||} 
 \hline
   Parameters: & n & m & $\hat{Q}$& $\hat{l}_c$& $E_{\textrm{max}}$/PeV \\ [0.8ex]
\hline \hline 
 Case \textit{1} & 1 & 0 & 7.4 & 1  & 3\\ 
 Case \textit{2} & 1 & 0 & 12.6 & 3  & 3 \\
 Case \textit{3} & 1 & 0 & 20.2 & 10  &  3\\[1.5ex]
 Case \textit{4} & 0.9 & 0 & 5.8 & 1  &  3\\
 Case \textit{4b} & 0.9 & 0 & 6.2 & 1.2  &  3\\
 Case \textit{5} & 0.9 & 0 & 10.4 & 3 &  3\\
 Case \textit{6} & 0.9 & 0 & 18.3 & 10 & 3\\ [1.5ex] 
 Case \textit{7} & 1 & 0.2 & 11 & 3 &  3\\ 
 Case \textit{8} & 1 & 0.4 & 9 & 3 &  3\\[1.5ex] 
 Case \textit{9} & 0.9 & 0.2 & 5.4 & 1.2 & 3\\
 Case \textit{10} & 0.9 & 0.4 & 4.3 & 1.2 & 3 \\[1.5ex] 
 Case \textit{11} & 1 & 0 & 10.2 & 3 & 1 \\
 Case \textit{12} & 1 & 0 & 15.2 & 3 & 10 \\ [1.5ex] 
 \hline
\end{tabular}
\medskip
\caption{Different model parameter combinations considered in this work. In all the cases, $\hat{B}=1$ and $q=2$.}
\label{tab:param}
\end{table}

\section{Gamma-ray emission}
\label{sec:gamma}

Using the CR density determined in \S~\ref{sec:crdensity}, in this section we calculate the gamma-ray spectrum due to the decay of neutral pions ($\pi^{0}$) produced in collisions between the CRs and the background gas, both assumed to be composed of protons. Following \cite{Aharonian_2004_VHE_book}, the gamma-ray flux per unit gamma-ray energy $E_\gamma$ received at Earth is given by 
\begin{linenomath}
\begin{equation}
\label{eqn: emission}
\begin{aligned}
\Phi_\gamma(E_\gamma)= &\,  \frac{2c}{d^2\kappa_{\pi}}   \int_{E_{\pi,\textrm{min}}}^{\infty}\,\mathrm{d}E_{\pi} \frac{\sigma_{pp}(E)}{\sqrt{E_{\pi}^2-(m_\pi c^2)^2}}
   \\ 
    & \times \, \int_{r_{\textrm{min}}}^{\infty}\,\mathrm{d}r \: r^2   n_{\textrm{gas}}(r) \frac{\mathrm{d}n_{\mathrm{CR}}}{\mathrm{d}E}(E
    ,r),
     \end{aligned}
\end{equation}
\end{linenomath}
where $E_{\pi,\textrm{min}}=E_\gamma+m_\pi^2c^4/4E_\gamma$, $E=m_p c^2+\frac{E_\pi}{\kappa_\pi}$, $E_{\pi}$ is the energy of the neutral pion produced in the collisions, $m_p$ and $m_{\pi}$ are the proton and pion masses, $\kappa_\pi\approx 0.17$ is the mean fraction of the CR kinetic energy transferred to neutral pions in the collisions \citep{Aharonian_2004_VHE_book}, $d \approx 8\,\mathrm{kpc}$ is the distance between Sgr~A* and the Earth, and $r_{\textrm{min}}$ is the minimum radius considered for the emissivity calculation. 
Unless stated otherwise, we take $r_{\textrm{min}}=10^{13}$ cm $\approx 10\,R_g$ in all our calculations. The cross-section for $\pi^0$ production through proton-proton collisions, in a proton reference frame, is given by \citep{Aharonian_2004_VHE_book}
\begin{linenomath}
\begin{equation}
        \label{eqn: cross}
\sigma_{pp}(E)\approx 4 \times 10^{-26}\,\left(1+0.04\,\ln\frac{E
-m_pc^2}{\mathrm{TeV}}\right) \,\text{cm}^2.
\end{equation}
\end{linenomath}
Since we are interested in very high energy protons, with $E\gtrsim 1\,\mathrm{TeV}$, we neglect the proton and pion masses in all calculations.
\begin{linenomath}
\begin{figure}[ht]
    \centering
    \includegraphics[width=0.47\textwidth]{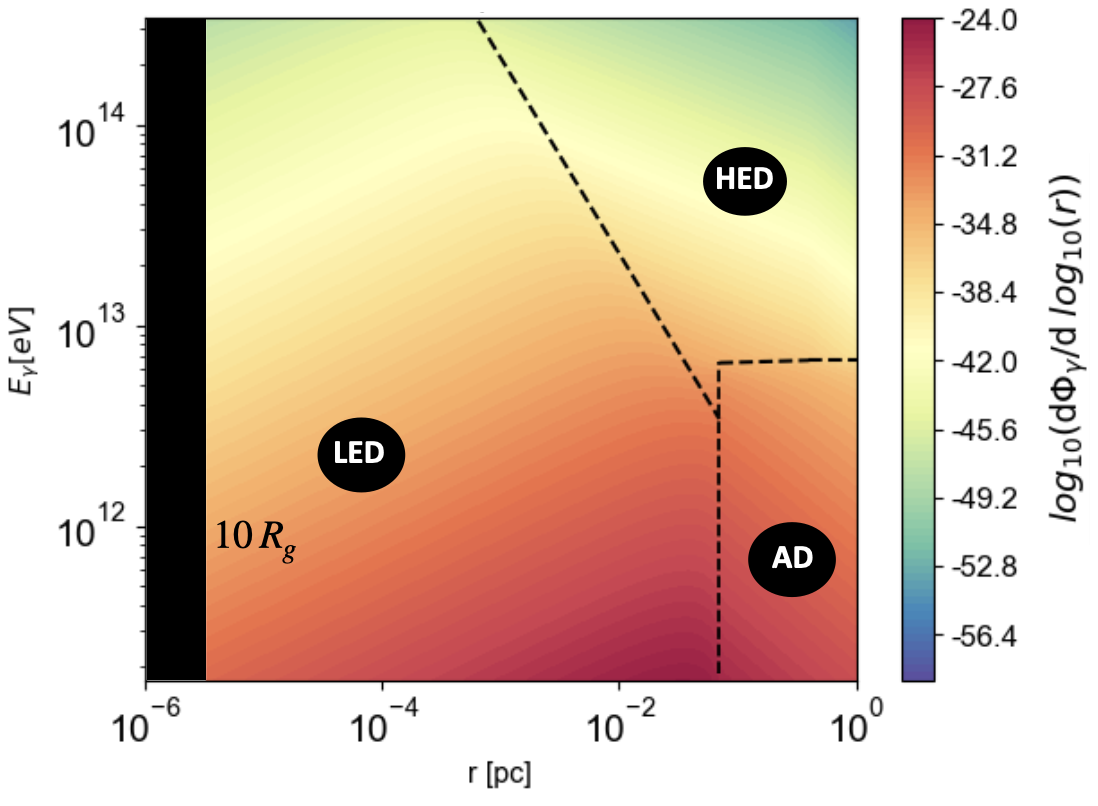}
     \caption{Differential gamma-ray flux per logarithmic interval of $r$, $d\Phi_\gamma/d\log_{10}(r)$, as a function of $E_\gamma$ and $r$, for Case \textit{2} in Table \ref{tab:param}}
    \label{fig: Dif_emiss}
\end{figure}
\end{linenomath}
\subsection{Behavior of the gamma-ray spectrum}
\label{sec:qualitative}

\begin{figure}[ht]
    \centering
    \includegraphics[width=0.47\textwidth]{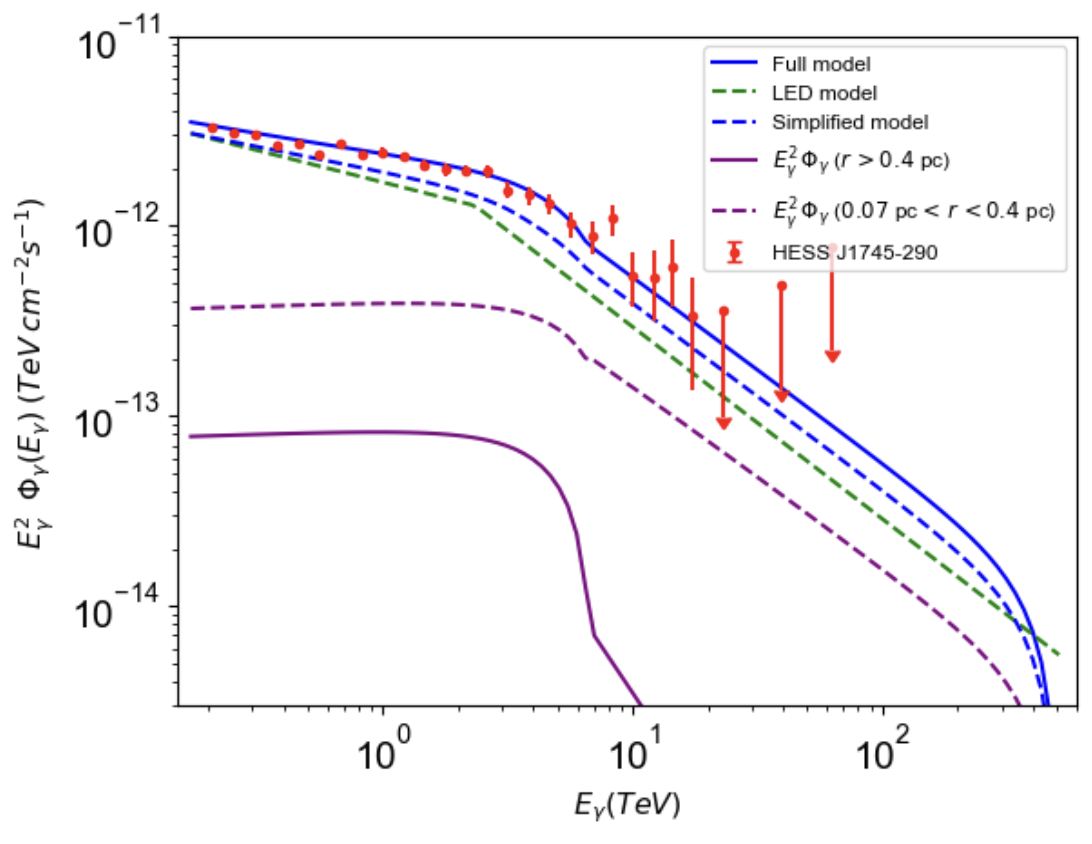}
     \caption{Full and approximate models for the gamma-ray spectrum of HESS~J1745$-$290 with the parameters of Case \textit{2} of Table 1. The solid blue line represents the total gamma-ray emission calculated from the full model. The dashed blue line is the ``LED model'', considering only the emission from CRs in the LED regime. The dashed green line shows our simplified model, given by Eqs.~(\ref{eqn: appx1}) and (\ref{eqn: appx2}). The solid and dashed purple lines represent the contributions from CRs at $r > 0.4$ pc and $0.07\,\mathrm{pc}<r<0.4\,\mathrm{pc}$, respectively. The red dots correspond to the HESS measurements 
     \citep{Abramowski_2016}.} 
    \label{fig:Emiss1}
\end{figure}

\begin{figure}[h]
    \centering
    \includegraphics[width=0.47\textwidth]{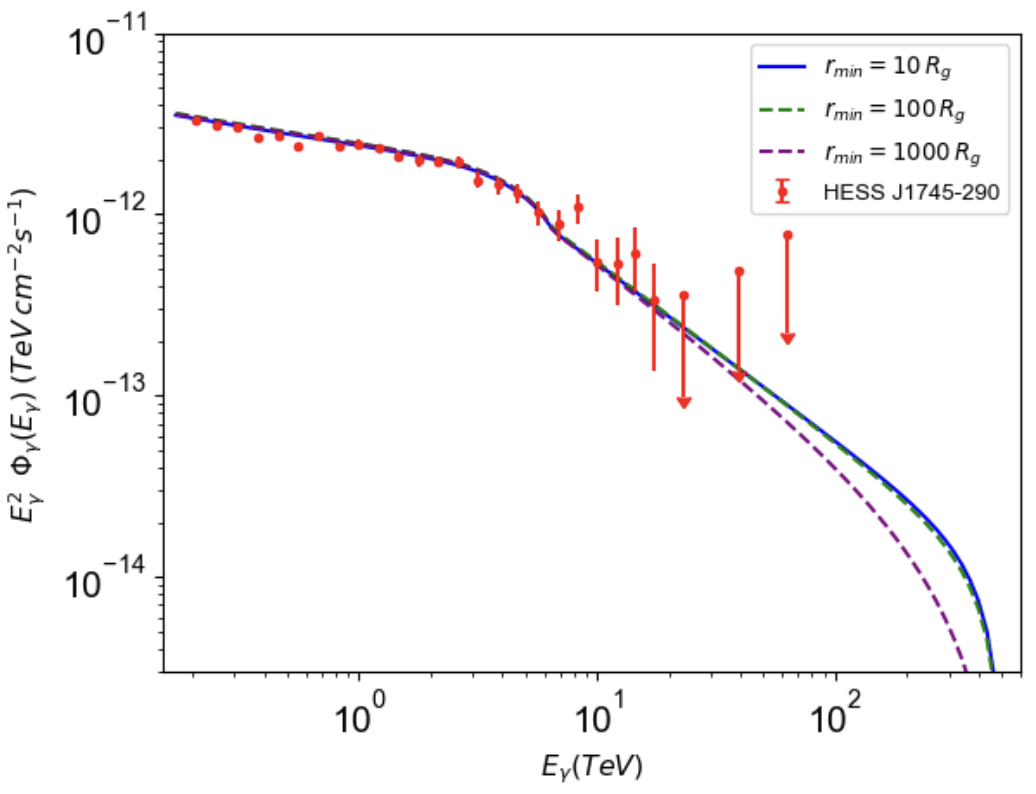}
     \caption{Gamma-ray spectra for Case \textit{2} of Table~\ref{tab:param}, for different minimum radii $r_{\mathrm{min}}$ from which the gamma-ray emission is integrated (see Eq.~(\ref{eqn: emission})). The red dots correspond to the H.E.S.S. measurements of the source HESS~J1745$-$290 \citep{Abramowski_2016}.} 
    \label{fig:Emiss_rmin}
\end{figure}

Figure ~\ref{fig: Dif_emiss} shows the differential gamma-ray flux as a function of $r$ and $E_\gamma$ for Case \textit{2} of Table~\ref{tab:param}. This flux clearly in correspondence with the CR density shown in Fig.~\ref{fig:CR_dens}, considering the approximate energy rescaling $E_\gamma\approx\kappa_\pi E$. Since $\mathrm{d}n_{\mathrm{CR}}/\mathrm{d}E$ drops off more quickly as a function of $r$ in the HED and AD regimes than in the LED regime, the gamma-ray flux turns out to be dominated by the CRs in the LED regime for any value of $E_\gamma$, with a maximum around the transition from the LED to the HED or AD regimes. This means that, in our model, the emission is dominated by $r\sim 0.07$ pc for $E_\gamma \lesssim 6$ TeV, and by $3\times 10^{-4}$ pc $\lesssim r\lesssim 0.07$ pc (equivalent to $10^3 R_g\lesssim r\lesssim 2\times 10^5 R_g$) for 6 TeV $\lesssim E_\gamma \lesssim 200$ TeV. In other words, the gamma-ray emission should only be significant in the range $10^3 R_g < r < 0.07$ pc. This can be seen from Fig. \ref{fig:Emiss1}, which shows the total emission spectrum for Case 2 (blue line) and compares it with the contributions from $r > 0.4$ pc (solid purple line) and from 0.07 pc $<r< 0.4$ pc (dashed purple line). We see that the emissions from these two ranges of radii only account for at most 
$\sim 20-30\%$ of the total emission at all the energies considered. Furthermore, one can also see from Fig.~\ref{fig:Emiss_rmin} that the resulting spectrum (up to $\sim 100$ TeV) for Case 2 is quite insensitive to the choice of $r_{\textrm{min}}$ as long as this parameter is in the range $\sim (10-10^3)\, R_g$, confirming that most of the emission comes from $r \gtrsim 10^3 R_g$. 

From Fig. \ref{fig: Dif_emiss} we also see that, for CR energies $E>E_C$, the LED region is constrained to progressively smaller radii $r_C(E)$ as $E$ increases, thus progressively decreasing the effective emitting volume. This causes a break in the gamma-ray spectrum (obtained by integrating the flux over the whole volume out to $r\rightarrow \infty$ and shown as the solid blue line in Fig.~\ref{fig:Emiss1}) at $E_{\gamma,b}\sim\kappa_\pi E_C$, beyond which it decreases more steeply, as in the observed spectrum of HESS~J1745$-$290.

The hypothesis that the gamma-ray spectrum is essentially determined by the shape of the LED region in Fig. \ref{fig:CR_dens} is confirmed by recalculating $\Phi(E_\gamma)$ considering only the gamma-ray emission from CRs in this regime (i.~e., setting $\mathrm{d}n_{\mathrm{CR}}/\mathrm{d}E = 0$ in the HED and AD regimes). This calculation is shown by the dashed blue line in Fig. \ref{fig:Emiss1}, which shows a similar broken power-law behavior as in the full calculation represented by the solid blue line. Thus, the predicted emission in our model is indeed determined by the CRs in the LED regime, and its broken power-law behavior is a consequence of the energy-dependent transition in the CR diffusion regime at $r_C(E)$ for $E > E_C$.

\subsection{Simplified model}
\label{sec:simple}
In order to understand how the main features of the HESS~J1745$-$290 spectrum restrict our model parameters, it is useful to provide approximate analytical expressions for our calculated gamma-ray emission. Since this emission is dominated by CRs in the LED regime, it can be approximated by considering only the contribution from that regime. For further simplicity, we approximate the CR density in the LED regime by Eq.~(\ref{Eq: CR dens1}) (neglecting subdominant terms that ensure continuity with the other regimes), ignore the (weak) energy-dependence of $\sigma_{pp}$ by fixing $E=1$ TeV in Eq.~(\ref{eqn: cross}), and integrate Eq.~(\ref{eqn: emission}) from $E_\pi=E_{\pi,\mathrm{min}}=E_\gamma$ to $E_{\pi,\mathrm{max}}=+\infty$ and from $r=0$ to $r=0.07\,\mathrm{pc}$ for $E_\gamma<E_{\gamma,b}$ and to $r=r_C(E)$ for $E_\gamma>E_{\gamma,b}$. (We note that the integral is dominated by the largest radii and the smallest energies integrated over.) This way, we obtain relatively simple expressions for $\Phi_\gamma(E_\gamma)$, which preserve the main features of our model, while showing the effect of the model parameters on the emitted spectrum. For $E_\gamma \lesssim E_{\gamma,b}$, 
\begin{linenomath}
\begin{equation}
    \label{eqn: appx1}
\begin{aligned}
  \Phi_\gamma(E_\gamma)\approx  & \,1.3 \times  10^{-9}  (2.1\times 10^4)^{-\frac{n}{3}} \, \frac{f(q,E_{\textrm{max}}) \: \hat{Q} \: \hat{l}_c^{-\frac{2}{3}}\: \hat{B}^{\frac{1}{3}}}{1-\frac{1}{9}(n+2m)^2} \\
  & \times \,  \frac{\kappa_{\pi}^{\,q-\frac{2}{3}}}{q+\frac{1}{3}}\, \left(\frac{E_\gamma}{ \mathrm{TeV}}\right)^{-q-\frac{1}{3}}\:\mathrm{TeV}^{-1}\:\mathrm{cm}^{-2}\:\mathrm{s}^{-1},
\end{aligned} 
\end{equation}
\end{linenomath}
while, for $E_\gamma \gtrsim E_{\gamma,b}$,
\begin{linenomath}
\begin{equation}
    \label{eqn: appx2}
    \begin{aligned}
    \Phi_\gamma(E_\gamma)\approx  &\, 1.5\times  10^{-10} \: 6.2^{-\frac{m}{n-m}}  \, (2.1\times 10^4)^{-\frac{m(2-n)}{n-m}} \, \frac{f(q,E_{\textrm{max}}) \, \hat{Q}}{1-\frac{1}{9}(n+2m)^2}\\
  &  \times \, \frac{\kappa_{\pi}^{q+\frac{1-n}{n-m}}}{q+\frac{1-m}{n-m}} \, \hat{l}_c^{\frac{1-n}{n-m}}\, \hat{B}^{\frac{1-m}{n-m}} \left(\frac{E_\gamma}{ \mathrm{TeV}}\right)^{-q-\frac{1-m}{n-m}}\:\mathrm{TeV}^{-1}\:\mathrm{cm}^{-2}\:\mathrm{s}^{-1},
    \end{aligned}
\end{equation}
\end{linenomath}
both of which are represented by the dashed green line in Fig.~\ref{fig:Emiss1}.
 
In order to reproduce the observations of HESS~J1745$-$290, the broken power-law predicted by our model needs to reproduce the values for its low and high-energy indices, $\alpha_{LE}$ and $\alpha_{HE}$ (defined as $\alpha\equiv-\mathrm{d}\log\Phi_\gamma/\mathrm{d}\log E_\gamma$ in the respective regimes), as well as the energy of the spectral break, $E_{\gamma,b}$.

Assuming $q\approx 2$, as suggested by the spectrum in the CMZ (see \S~\ref{sec:approxeffect}), our model yields $\alpha_{LE}=q+1/3\approx 2.3$, in good agreement with the observations \citep{Abramowski_2016}, as can be seen from Fig.~\ref{fig:Emiss1}. At high energies, assuming also $n\approx 1$, as suggested by the previous MHD simulations, our model yields $\alpha_{HE}=q+(1-m)/(n-m)\approx 3$, fairly independent of the value of $m$. 
This result also appears to be consistent with the H.E.S.S. data (as seen in Fig. \ref{fig:Emiss1}), although this part of the spectrum is less well constrained by the observations. 

If $E_{\gamma,b}\approx 3$ TeV, as suggested by the observed spectrum, the critical energy $E_C= E_{\gamma,b}/\kappa_\pi\approx 20$ TeV. Comparing to Eq.~(\ref{eqn: E_crit}) with $\hat{B}\approx 1$ and $n\approx 0.9-1$, as inferred from the MHD simulations of the Sgr A* accretion flow, this strongly constrains $\hat{l}_c \approx 1-3$.

\subsection{Constraints from our full model}
\label{sec:full}

\begin{figure}[h]
    \centering
    \includegraphics[width=0.45\textwidth]{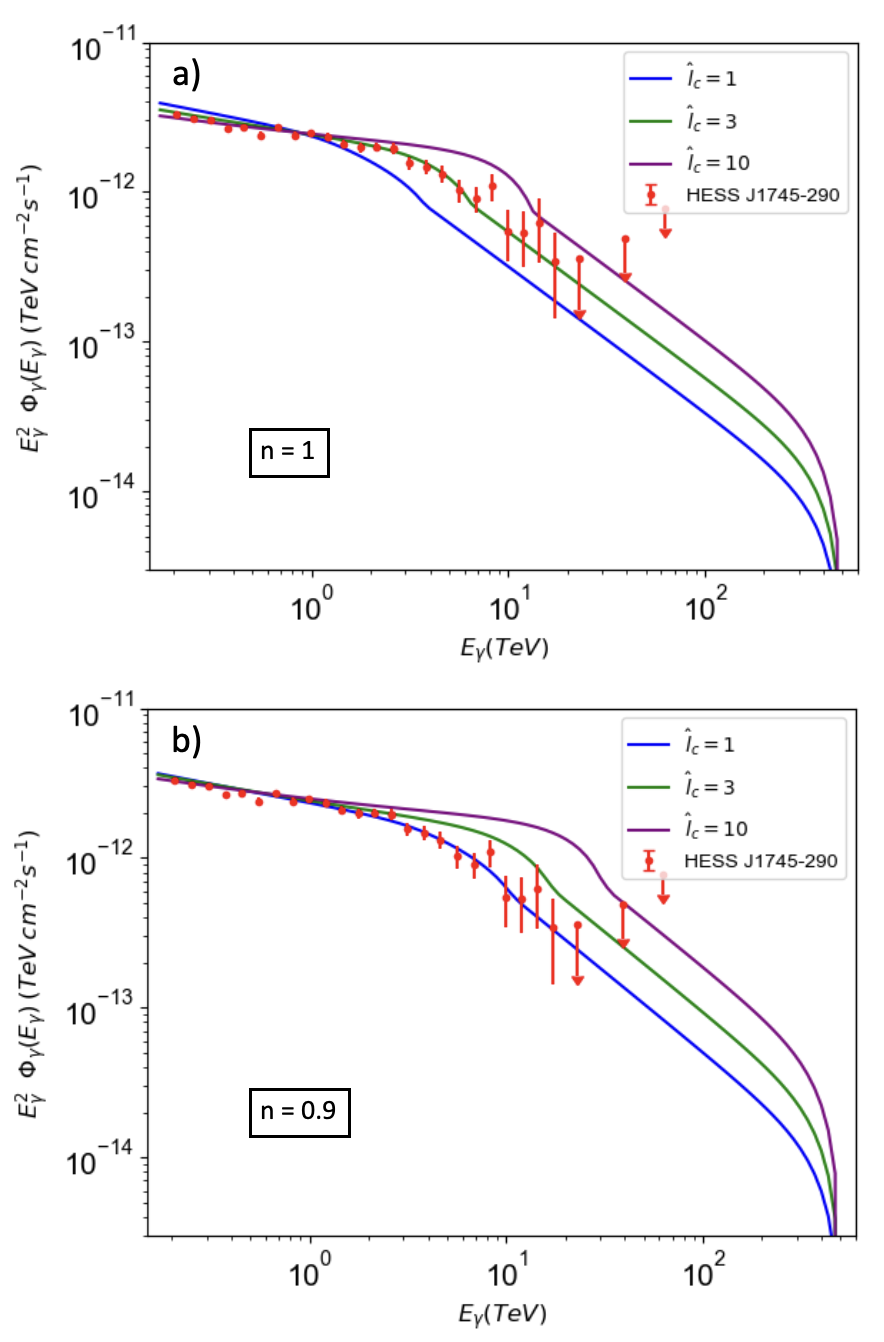}
    \caption{Gamma-ray spectrum for different choices of the parameters $n$ and $\hat l_c$. Panel~\textit{a}: Cases 1 to 3 of Table~\ref{tab:param}, all with $n=1$. Panel~\textit{b}: Cases 4 to 6 (excluding Case 4$b$), with $n=0.9$. The three curves on each panel correspond to different values of $\hat l_c$, with $\hat Q$ chosen so as to produce the same flux at $E_\gamma=1\,\mathrm{TeV}$ for all cases. The red dots are obtained from  \cite{Abramowski_2016}.}   
    \label{fig:E(lc)}
\end{figure}

In this section, we use our full model to show how our parameters are further constrained by the main features of the HESS~J1745$-$290 spectrum. 

The restrictions on $\hat{l}_c$ can be seen from Figure \ref{fig:E(lc)}.
It confirms that, for $n=1$ and $n=0.9$, the data favor the values $\hat{l}_c=3$ and $\hat l_c=1$, respectively. 
As expected, the effect of increasing $\hat l_c$ 
is to proportionally increase $E_C$ and, therefore, the energy of the spectral break, $E_{\gamma,b}$, as seen from Equation~(\ref{eqn: E_crit}). This equation implies that cases with different $\hat{l}_c$ can, or course, produce the same $E_{\gamma,b} \approx 6$ TeV as long as $\hat{B}$ and $n$ are adjusted accordingly. However, as argued above, the fact that the MHD simulations favor $\hat{B}=1$ and $n=0.9-1$ strongly constrains $\hat{l}_c$ to the rather small range $\hat{l}_c\approx 1-3$.

\begin{figure}[ht]
    \centering
    \includegraphics[width=0.47\textwidth]{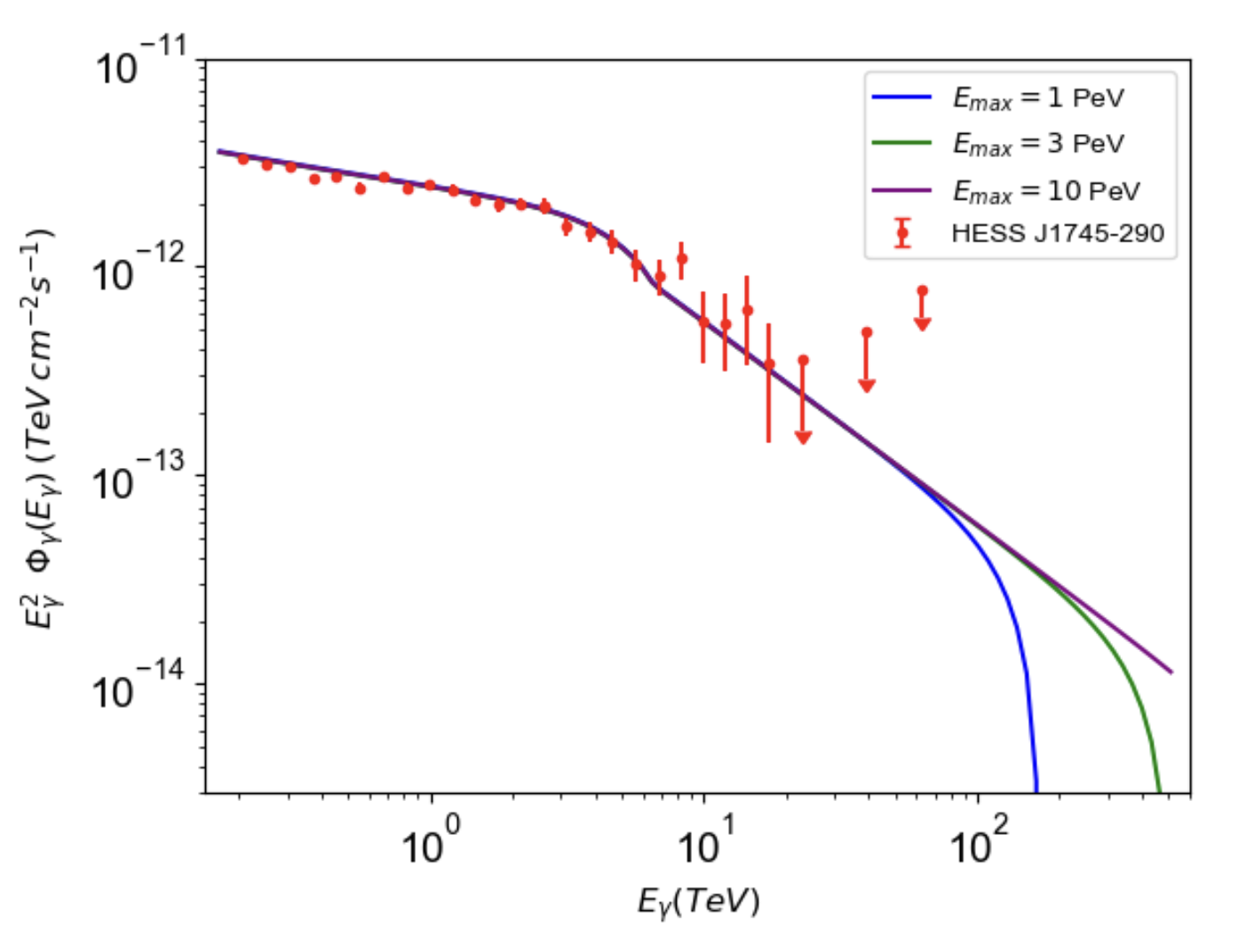}
     \caption{Gamma-ray spectrum for different maximum energies $E_{\textrm{max}}=1$, 3 and 10 PeV (Cases \textit{11}, \textit{2} and \textit{12}, respectively), all with the same parameter values $n=1$, $m=0$, $q=2$, $\hat{B}=1$, $\hat{l}_c=3$, and slightly different values $\hat{Q}$, chosen so that the three spectra visually fit the H.E.S.S. data.}
    \label{fig: E(max)}
\end{figure}

%
The effect of the cutoff energy of the CR injection spectrum, $E_{\textrm{max}}$, 
is shown in 
Fig. \ref{fig: E(max)}. As expected, it causes the gamma-ray spectrum to cut off around $E_\gamma\sim\kappa_\pi E_{\mathrm{max}}$. 

\begin{figure}
    \centering
    \includegraphics[width=0.45\textwidth]{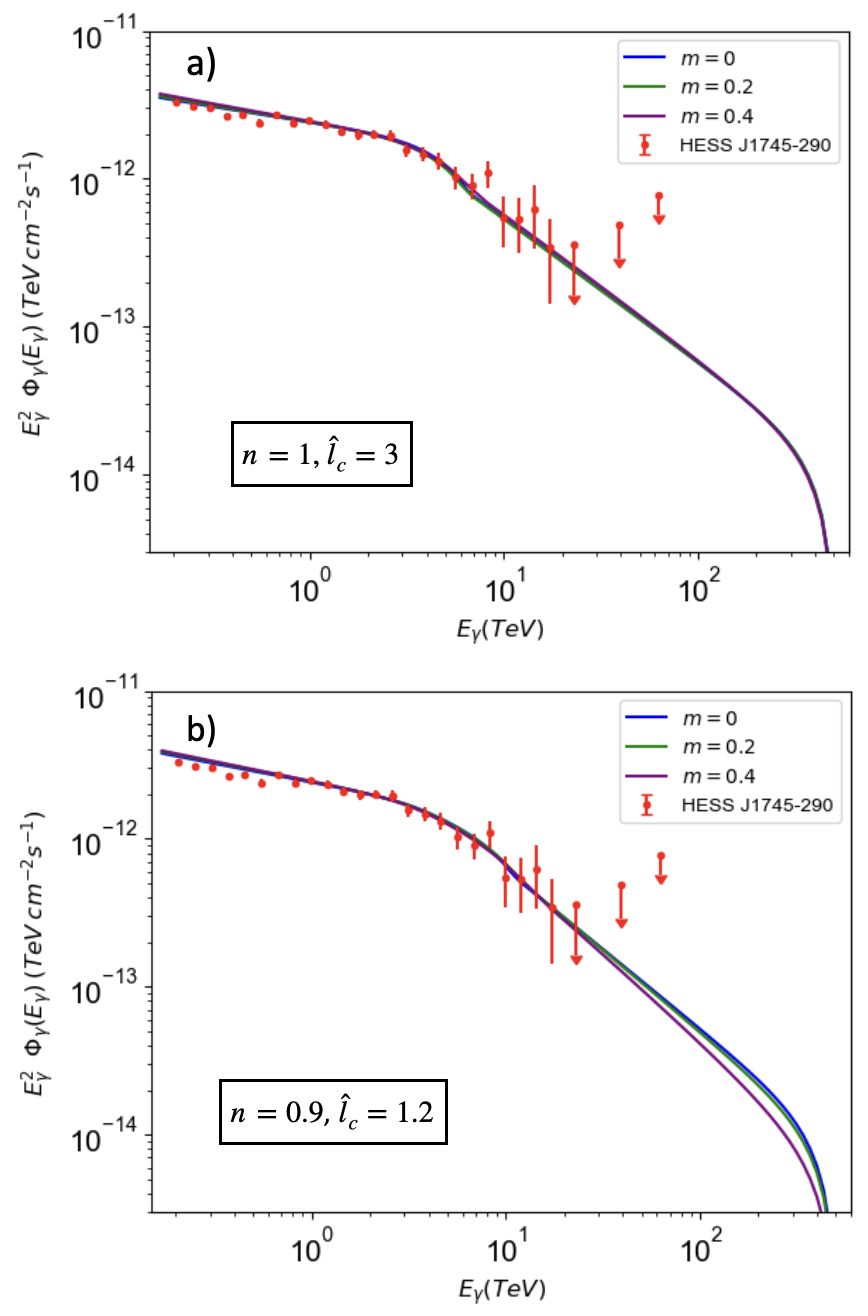}
     \caption{Dependence of the gamma-ray spectrum on the exponent $m$ of the coherence length profile. Panel $a$: $n=1,$ $\hat{l}_c=3,$ $m=0, 0.2, 0.4$ (Cases \textit{2}, \textit{7} and \textit{8} in Table \ref{tab:param}). Panel $b$: $n=0.9,$ $\hat{l}_c\approx 1,$ $m=0, 0.2, 0.4$ (Cases \textit{4}, \textit{9} and \textit{10}).} 
     \label{fig: E(m)}
\end{figure}  

In \S~\ref{sec:simple}, we used approximate expressions for $\Phi_{\gamma}(E_{\gamma})$ (Eqs.~\ref{eqn: appx1} and \ref{eqn: appx2}) to argue that the exponent $m$ of the radial profile of the coherence length $l_c$ should not significantly affect the spectral index of the gamma-ray spectrum at low or high energies ($\alpha_{LE}$ and $\alpha_{HE}$, respectively) or the gamma-ray energy that separates these two power-law regimes ($E_{\gamma} \sim 6$ TeV). This is largely 
confirmed by Fig.~\ref{fig: E(m)}$a$, which compares spectra for different values of $m$ in the case $n=1$. 
However, Eq.~(\ref{eqn: appx2}) implies that, in the case $n=0.9$, increasing the parameter $m$ can weakly increase the index $\alpha_{HE}$ of the high-energy emission, making this part of the spectrum steeper. This is indeed what is seen in Fig.~\ref{fig: E(m)}$b$, where
only the case $m=0.4$ shows a 
noticeably steeper spectrum. However, we will see in \S~\ref{sec:consistency} that the self-consistency of our model requires $m$ to be small ($\lesssim 0.3$), so we do not expect any significant effect of the parameter $m$ as long as this restriction is met.

We have thus shown that models with $\hat{B} = 1$, $q=2$, and $E_{\textrm{max}} \gtrsim 1$ PeV can reproduce well the main features of the spectrum of HESS~J1745$-$290, including its break at $E_{\gamma,b}\approx 6$ TeV. This happens for a narrow range of parameters, spanned by the 
cases with $n\approx 1$, $\hat{l}_c\approx 3$, and $\hat{Q}\approx 13$ on the one hand, and $n\approx 0.9$, $\hat{l}_c\approx 1$, and $\hat{Q}\approx 6$ on the other. Remarkably, these values of $q$, $\hat{Q}$ and $E_{\textrm{max}}$ are in good agreement with the estimates from the diffuse CMZ emission, in particular those obtained by \cite{Abramowski_2016}. Regarding $m$, its effect on the spectrum is rather weak, and it only appears at $E_{\gamma}\gtrsim E_{\gamma,b}$. These results, however, establish a strong restriction on the largely unknown parameter $\hat{l}_c$, which needs to be in the range $\hat{l}_c \approx 1-3$ (equivalent to a coherence length of $l_c \approx (1-3)\times 10^{14}$ cm). 

\section{Consistency}
\label{sec:consistency}

In this section we check the consistency of several simplifying assumptions made in the calculation of the gamma-ray emission.   

\subsection{Gas density profile $n_{\textrm{gas}}(r)$ for 0.07 pc $< r <$ 0.4 pc.}
\label{sec:densityII}
According to Fig. 11 of \citealt{Ressler_2018}, the $n_{\textrm{gas}}(r)$ profile at $0.07 \textrm{ pc}\lesssim r \lesssim 0.4 \textrm{ pc}$ is steeper than the $n_{\textrm{gas}}(r) \propto r^{-1}$ profile assumed in our calculations (Eq. \ref{eq. gas dens}). This means that our calculated gamma-ray emission is somewhat larger than what we would obtain if a more accurate profile were assumed.

Here we show that using Eq. \ref{eq. gas dens} to model $n_{\textrm{gas}}(r)$ in the range $0.07 \textrm{ pc}< r < 0.4 \textrm{ pc}$ only leads to a $\sim 20-30\%$ overestimate in the emission and, therefore, does not affect significantly our results. This is shown in Fig. \ref{fig:Emiss1}, where the purple dashed line represents the emission from $0.07 \textrm{ pc}< r < 0.4 \textrm{ pc}$ in our Case 2. We see that this emission is $\lesssim 20-30 \%$ of the total emission at all the energies of interest. Thus, assuming $n_{\textrm{gas}}(r) \propto r^{-1}$ at all radii $r<0.4$ pc should not affect significantly our main results.

\subsection{Cosmic ray transport for radii $r>0.4$ pc}\label{sec:transport>0.4}

Even though CR transport at radii $r>0.4$ pc is expected to be a combination of advection and diffusion, in our calculations we have simply assumed that transport at $r>0.4$ pc is the same as at $r=0.4$ pc. This should lead to an overestimate of the CR density in this region (since the assumed transport process may not be the most efficient one) and, therefore, provides an upper limit to its corresponding emission from $r>0.4$ pc. In Fig. \ref{fig:Emiss1}, the purple line represents this upper limit to the gamma-ray emission from $r>0.4$ pc. We see that this emission is $\lesssim 3 \%$ of the total emission for all the energies of interest, so this overestimate should not affect the accuracy of our results.

\subsection{Validity of diffusion at $r < 0.07$ pc}
\label{sec:minradius}

In our calculations, we assume that diffusion is the dominant process for CR transport between $r=r_{\mathrm{min}}\sim 10\,R_g$ (where we expect the CRs to be injected) and $r = 0.07$ pc. In particular, at $r\sim 10\,R_g$, diffusion in the LED regime is assumed for CRs of all energies (as seen in the two cases (Case 2 and Case 4$b$) shown in Fig. \ref{fig:CR_dens}). However, diffusive transport is a valid approximation for CR propagation only when $r\gg r_{\lambda}$, where $r_{\lambda}$ is defined as the radius where the CRs' mean free path equals the radius itself, that is,
\begin{linenomath}
\begin{equation}
    \label{eq. rmin}
    r_{\lambda}=\lambda_{\textrm{mfp}}(r_{\lambda}).
\end{equation}
\end{linenomath}
From Eq.~(\ref{eq:D1}), we can obtain the expression for the mean free path of the CRs in the LED regime and show that $r_{\lambda}$ satisfies
\begin{linenomath}
\begin{equation}
 \begin{aligned}
 \left(\frac{r_{\lambda}}{10R_g}\right)^{1-\frac{1}{3}(n+2m)} = & \, 6 \times 10^{-2} \, (2.1\times 10^4)^{-2m/3} \, \hat{l}_c^{2/3} \, \hat{B}^{-1/3}\,\left(\frac{E}{1\text{TeV}}\right)^{1/3}. \\
 \end{aligned}
\label{eq:rlambda}
\end{equation}
\end{linenomath}

Since the right hand side of Eq.~(\ref{eq:rlambda}) is a growing function of $E$, its upper limit is obtained by evaluating it at the maximum CR energy considered in this work, $E=10$ PeV. This way we see that $r_{\lambda}/(10\,R_g)$ is always $\lesssim 2.7$, implying that assuming diffusion in the LED regime  is a valid approximation for $r$ larger than $\sim 30 R_g$, which is right outside the region where CRs are injected in our model.

We note, however, that applying Eq.~(\ref{eq:rlambda}) to values of $r_{\lambda}$ as small as $r\sim 10 \,R_g$ may not be entirely consistent with the fact that Eq.~(\ref{eqn: lc}) predicts the existence of a radius $r_{\textrm{eq}}$ defined by
\begin{linenomath}
\begin{equation}
\label{eq:req}
    r_{\textrm{eq}} = l_c(r_{\textrm{eq}}),
\end{equation}
\end{linenomath}
such that, at $r\lesssim r_{\textrm{eq}}$, Eq.~(\ref{eqn: lc}) is no longer a valid description for $l_c(r)$ (because $l_c(r)$ cannot be larger than $r$). Also, at $r<r_{\textrm{eq}}$, the diffusion process should behave more as a 1D process, in the sense that the propagation occurs along a nearly homogeneous background magnetic field. Therefore, in obtaining Eq.~(\ref{eq:rlambda}), we are implicitly making the simplifying assumption that the dependence of the CR mean free path on $r$ at $r<r_{\textrm{eq}}$ is the same as for $r>r_{\textrm{eq}}$. Given that this simplification only applies to $r$ between $\sim 10R_g$ ($\sim 10^{13}$ cm) and $r_{\textrm{eq}}$ ($\lesssim 10^{14}$ cm for $\hat{l}_c \sim 1$), whereas most of the emission comes from large radii, we believe that it 
should not affect significantly the accuracy of our results.

\subsection{Neglecting advection for $r<0.07$ pc}\label{sec:negadvect}

Besides diffusion, advection can also contribute to CR transport due to the average radial gas velocity $v_{\textrm{gas}}$ at $r<0.07$ pc. \cite{Ressler_2018} show that for $0.01\,\text{pc}\lesssim r\lesssim 0.07\,\text{pc}$ there is a stagnation region where the mass accretion rate averaged over the whole solid angle approaches $\dot{M} \approx 0$, indicating that $v_{\textrm{gas}}$ should have a small effect on the CR transport in that region. In addition, for $r\lesssim 0.01$ pc there is an inflow-dominated region where $\dot{M}$ is nearly constant. Taking $\dot{M}\approx  10^{-8} M_{\odot}$/year near the black hole \citep{Dexter_2020}, we obtain the expression for $v_{\textrm{gas}}$ given by Eq.~(\ref{eqn:vr}). In order to neglect the effect of $v_{\textrm{gas}}$ in the transport of the CRs, its magnitude has to be smaller than $v_{\textrm{diff}}$ for the lowest energy CRs, which diffuse in the LED regime for all radii $r < 0.07$ pc (as seen in Fig. \ref{fig:CR_dens}). Considering the definition of $v_{\textrm{diff}}$ from Eq.~(\ref{eq. vdiff}), we obtain:
\begin{linenomath}
\begin{equation}
\label{eqn: vdif 2}
\begin{aligned}
 v_{\textrm{diff}}\approx &   \,3\times 10^6 \: 0.52^{2m} \: 14.4^{n-1} \, \hat{l}_c^{2/3}\,\hat{B}^{-1/3} \\
 &   \times \left( \frac{E}{\text{TeV}}\right)^{1/3}\left(\frac{r}{0.01\text{pc}}\right)^{\frac{1}{3}(n+2m)-1}\text{cm/s},
\end{aligned}
\end{equation}
\end{linenomath}
from which we obtain that the ratio $v_{\textrm{diff}}/|v_{\textrm{gas}}|$ is
\begin{linenomath}
\begin{equation}
\label{eqn:ratio}
\begin{aligned}
\frac{v_{\textrm{diff}}}{|v_{\textrm{gas}}|}\approx &   \,90 \cdot 0.52^{2m} \: 14.4^{n-1} \hat{l}_c^{2/3}\hat{B}^{-1/3}\left( \frac{E}{\text{TeV}}\right)^{1/3}\\
 &   \times \left(\frac{r}{0.01\text{pc}}\right)^{\frac{1}{3}(n+2m)}.
\end{aligned}
\end{equation}
\end{linenomath}
Since we are focusing on emission with $E_{\gamma} \gtrsim 0.2$ TeV, the lowest CR energy that we consider is $E\sim 0.2/\kappa_{\pi}$ TeV $\approx 1$ TeV. Thus, since the ratio $v_{\textrm{diff}}/|v_{\textrm{gas}}|$ is an increasing function of $r$, neglecting advection requires $v_{\textrm{diff}}/|v_{\text{gas}}|\gtrsim 1$ for $E=1$ TeV and $r=10 R_g$.

\begin{figure}[ht]
    \centering
    \includegraphics[width=0.45\textwidth]{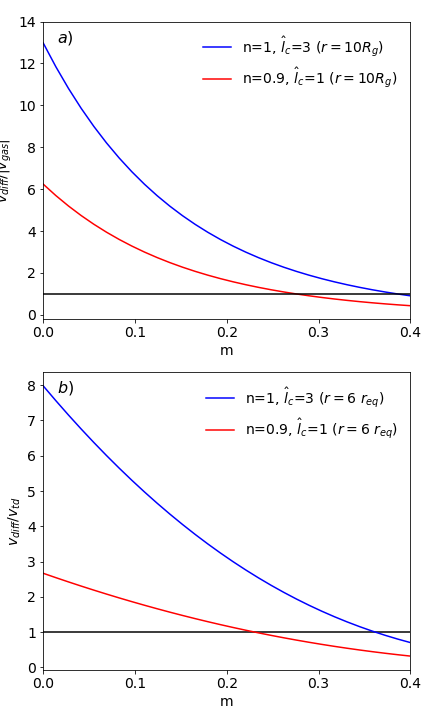}
 \caption{Panel $a$ shows $v_{\textrm{diff}}/|v_{\textrm{gas}}|$ (Eq.~(\ref{eqn:ratio})) as a function of $m$ for $E=1$ TeV and $r=10 R_g$. Panel $b$ shows $v_{\textrm{diff}}/v_{td}$ (Eq.~(\ref{eq:vdvtd})) as a function of $m$ for $E=1$ TeV and $r=6 r_{\mathrm{eq}}$. In both panels, the blue line represents the case $n=1$, $\hat{l}_c = 3$, and the red line corresponds to $n=0.9$, $\hat{l}_c = 1$, both with $\hat B=1$.}
    \label{fig:compare}
\end{figure}

Fig.~\ref{fig:compare}$a$ shows $v_{\textrm{diff}}/|v_{\textrm{gas}}|$ as a function of $m$ for $E=1$ TeV and $r=10 R_g$, for the two favored cases in this work: $n=1$, $\hat{l}_c = 3$ and $n=0.9$, $\hat{l}_c = 1$, shown in blue and red, respectively. We see that for $n=1$, $\hat{l}_c = 3$ neglecting advection requires $m \lesssim 0.4$, while for $n=0.9$, $\hat{l}_c = 1$ it requires $m \lesssim 0.3$. This result shows that the consistency of our calculations demands a weak dependence of $l_c$ on $r$, as we have assumed throughout this work.

Interestingly, the dependence of $v_{\textrm{diff}}/|v_{\textrm{gas}}|$ on $E$ also implies that, in our model, CRs of energies $E \ll $ 1 TeV are significantly affected by advection towards the black hole, in principle not being able to propagate beyond $r \sim 10 R_g$. This implies that a significant fraction of these particles should simply be accreted onto the black hole, with their gamma-ray emission for $E_{\gamma} \ll \kappa_{\pi}\times 1 \textrm{TeV}\sim 0.2$ TeV not being  captured by our model.  
 
\begin{figure}[ht]
\centering    \includegraphics[width=0.48\textwidth]{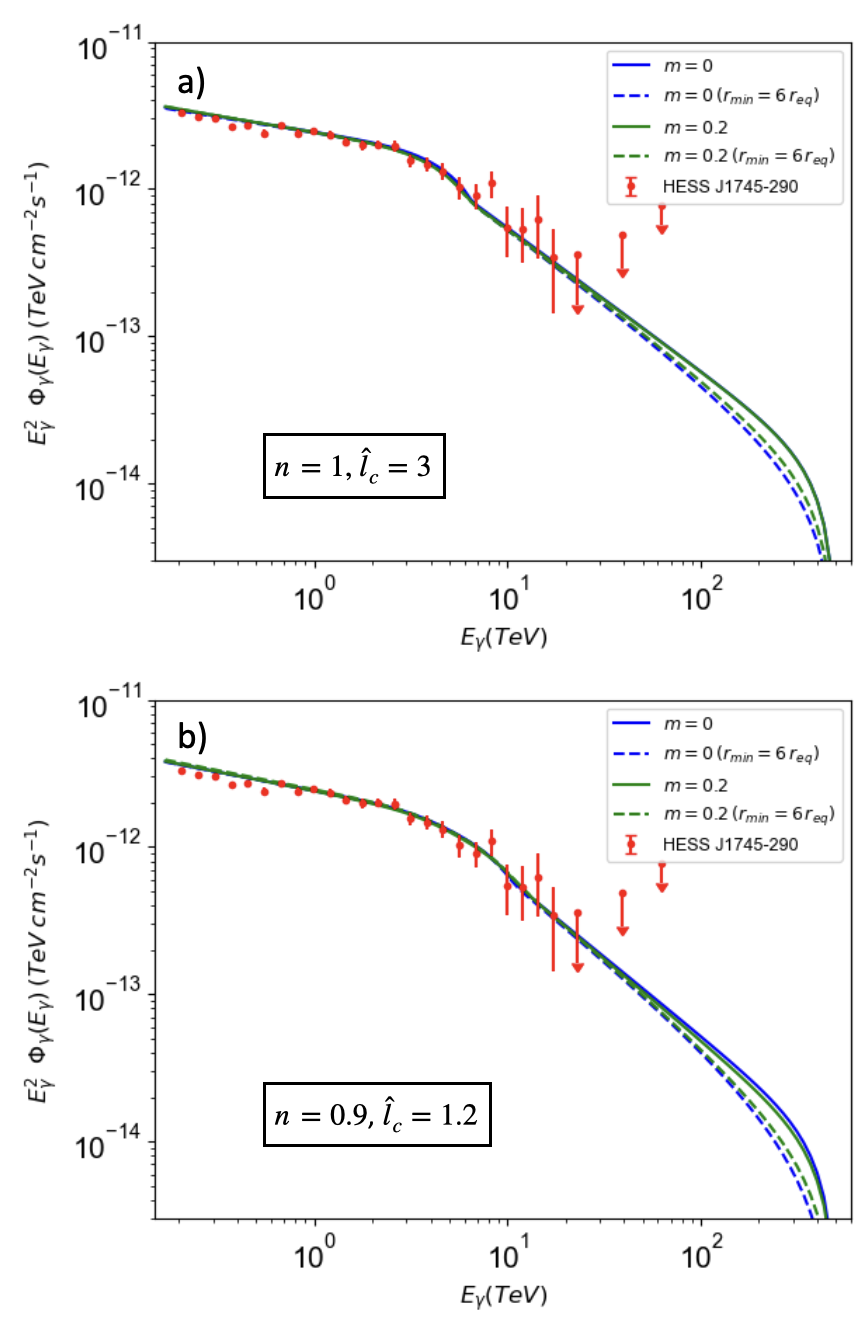}
\caption{
Effect of using different minimum radii $r_{\textrm{min}}$ on the gamma-ray spectra. The solid and dashed lines represent analogous cases using $r_{\textrm{min}}=10\,R_g$ and $r_{\textrm{min}}=6\,r_{eq}$, respectively. Panels $a$ and $b$ show 
the cases $n=1$, $\hat{l}_c=3$ and $n=0.9$, $\hat{l}_c=3$, respectively. In both panels we consider $m=0$ (blue lines) and $m=0.2$ (green lines).}
\label{fig: E(rmin)}
\end{figure}

Cosmic rays could also be advected by the random motions of the MHD turbulence in the accretion flow, which we neglected in our model. If we estimate this turbulent velocity by the Alfvén velocity of the gas, $v_A=B/\sqrt{4\pi n_{\mathrm{gas}} m_p},$ and its length-scale by $l_c$, the motion of the turbulence eddies should induce a CR diffusion characterized by a turbulent diffusion coefficient $D_t$ given by
\begin{linenomath}
\begin{equation}
\label{eq:dt}
    D_t\sim l_c v_A/3\text{,}
\end{equation}
\end{linenomath}
which allows us to estimate a turbulent diffusion velocity, $v_{td}$, as
\begin{linenomath}
\begin{equation}
\label{eq:vtd}
    v_{td}=D_t/r\text{.}
\end{equation}
\end{linenomath}
Combining Eqs.~(\ref{eqn: B}), (\ref{eqn: lc}), (\ref{eq. gas dens}), (\ref{eqn: vdif 2}), (\ref{eq:dt}), and (\ref{eq:vtd}), one can show that 
\begin{linenomath}
\begin{equation}
\label{eq:vdvtd}
\begin{aligned}
\frac{v_{\textrm{diff}}}{v_{td}} \approx & \, 120\,(1.9)^m(43200)^{n-1}\hat{l}_c^{-\frac{1}{3}}\hat{B}^{-\frac{4}{3}}\left(\frac{E}{\textrm{TeV}}\right)^{\frac{1}{3}}\\
& \times\left(\frac{r}{0.01 \,\textrm{pc}}\right)^{\frac{4}{3}n-\frac{1}{3}m-\frac{1}{2}},
\end{aligned}
\end{equation}
\end{linenomath}
which, for the values of $n$ and $m$ of interest ($n\approx 1$ and $m\ll 1$), is a growing function of $r$. In Fig. \ref{fig:compare}$b$ we show $v_{\textrm{diff}}/v_{td}$ as a function of $m$ in the case $r=6\, r_{\mathrm{eq}}$, where $r_{\textrm{eq}}$ is defined by Eq.~(\ref{eq:req}), and for the smallest relevant energy $E=1$ TeV. We can see that, in the cases of interest shown in the figure, 
$v_{\textrm{diff}}/v_{td} \gtrsim 1$ as long as $m\lesssim 0.3$, which also satisfies the conditions for neglecting the effect of $v_{\textrm{gas}}$. Thus, since $v_{\textrm{diff}}/v_{td}$ is a growing function of $r$, $v_{td}$ can be safely neglected for $r\gtrsim 6\, r_{\textrm{eq}}$. On the other hand, the effect of turbulent diffusion should be valid only for $r \gtrsim r_{\textrm{eq}}$. This means that there is a range of radii, $r_{\mathrm{eq}}\lesssim r\lesssim 6\,r_{\mathrm{eq}},$ in which turbulent diffusion can contribute significantly to CR transport, implying that the CR density $\mathrm{d}n_{\mathrm{CR}}/\mathrm{d}E$ in that region should be somewhat smaller than what is obtained in our model. In order to find an upper limit to this effect, 
Fig. \ref{fig: E(rmin)} compares gamma-ray spectra calculated with $r_{\mathrm{min}}=10\,R_g$ and with $r_{\mathrm{min}}=6\, r_{\mathrm{eq}}$, otherwise with the same parameters. 
We see that there are only small differences (at $E_{\gamma}\gtrsim 100$ TeV) between the calculations with $r_{\mathrm{min}}=R_g$ and $r_{\mathrm{min}}=6\, r_{\mathrm{eq}}$, implying that neglecting turbulent diffusion should not affect significantly the accuracy of our results.

This result also reinforces the idea that gamma-ray emission is mainly produced at large radii, either close to 0.07 pc for $E_{\gamma} \lesssim 6$ TeV or close to $r_C(E_{\gamma}/\kappa_{\pi})$ for $E_{\gamma} \gtrsim 6$ TeV, as shown in Fig \ref{fig: Dif_emiss}. 

\subsection{Cosmic ray, gas and magnetic  pressures}
\label{sec:pressure}

\begin{figure}
    \centering
    \includegraphics[width=0.47\textwidth]{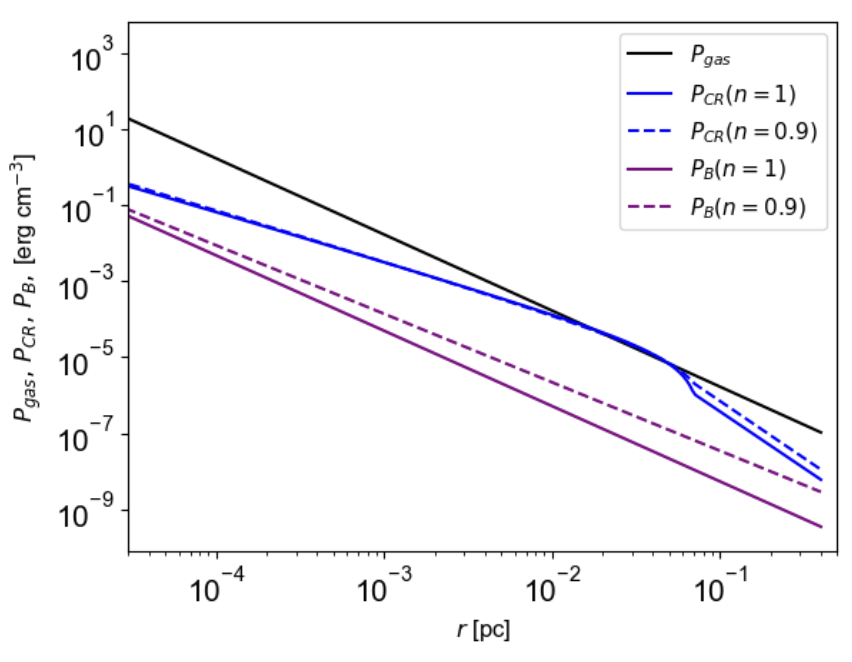}
     \caption{Cosmic-ray, gas and magnetic pressure comparison ($P_{\textrm{CR}}$, $P_{\textrm{gas}}$ and $P_B$, respectively). The values for the parameters used in the estimation of the CR pressure was done for Cases \textit{2} (solid lines) and \textit{4}$b$ (dashed lines) in Table \ref{tab:param}.}
    \label{fig: Pressure}
\end{figure}

Our calculations assume that the gas properties in the accretion flow are determined by the hydrodynamic evolution of the wind of $\sim 30$ WR stars that feed Sgr~A*, as in the simulations of \cite{Ressler_2018,Ressler_2020_small_B} on which our results are based. However, the presence of CRs diffusing out from the central black hole can be dynamically important in this evolution if the CR pressure becomes comparable to the gas pressure. Figure \ref{fig: Pressure} shows these two pressures as functions of $r$ for the Cases \textit{2} and \textit{4}$b$ in Table \ref{tab:param} (corresponding to our fiducial cases with $n=1$ and $n=0.9$ shown in Fig. \ref{fig:CR_dens}). The CR pressure is calculated as 
\begin{linenomath}
\begin{equation}
    P_{\mathrm{CR}}(r)=\frac{1}{3} \int_{E_{min}}^{E_{\textrm{max}}}\: \mathrm{d}E\:E \: \frac{\mathrm{d}n_{\mathrm{CR}}(E,r)}{\mathrm{d}E},
\end{equation}
\end{linenomath}
where $E_{min}=1$ TeV (since the CR density should be significantly suppressed for $E \ll 1$ TeV, as shown in \S~\ref{sec:negadvect}), and $E_{\textrm{max}}=3$ PeV, while the gas pressure is calculated as 
\begin{linenomath}
\begin{equation}
P_{\mathrm{gas}}(r)=2 \, n_{\textrm{gas}}(r) \, k_B \,T_{\textrm{gas}}(r),
\end{equation}
\end{linenomath}
where $k_B$ is the Boltzmann constant, $n_{\textrm{gas}}(r)$ is the gas density specified in Eq.~(\ref{eq. gas dens}), and $T_{\textrm{gas}}(r)$ is the angle- and time-averaged temperature of the gas obtained from Fig. 11 of \cite{Ressler_2018},\footnote{The time averages presented in Fig. 11 of \cite{Ressler_2018} are performed over the 100 years previous to present day.} as
\begin{linenomath}
\begin{equation}
    T_{\textrm{gas}}(r)=1.5\times 10^{7}  \left(\frac{r}{0.4 \text{pc}}\right)^{-1} \: \text{K}.
\end{equation}
\end{linenomath}

For completeness, we have also added the magnetic pressure profile for Cases 2 and 4$b$. We see that the gas pressure is dominant at almost all radii, with the CR pressure becoming relevant at $r\sim 0.07$ pc, around the inner boundary of the region where the WR star winds feed the accretion flow ($\sim 0.1-1$ pc). The fact that $P_{\mathrm{CR}}$ and $P_{\mathrm{gas}}$ are comparable at $r\sim 0.07$ pc may somewhat modify the evolution of the accreting gas near the feeding region. However, regarding the hydrodynamic properties of the gas, we use the fact that $P_{\mathrm{CR}}$ should only change the gas pressure by a factor $\sim 2$ near $r\sim 0.07$ pc, and thus
assume that the gas dynamics at those radii is fairly well described by hydrodynamic considerations. 

\section{Diffusion time-scales}\label{sec:Discussion}

The time-scale for diffusion from the central black hole to a certain radius $r$ can be estimated as $t_{\mathrm{diff}}\sim r^2/D(r)$ (valid as long as $D(r)$ grows more slowly than $r^2$). We have seen that, for CR energies $E<E_C\sim 20\,\mathrm{TeV}$, most of the emission occurs close to the radius $r\approx 0.07\,\mathrm{pc}$ where the transition from the LED to the advection regime occurs. Therefore, the time for CRs to diffuse out and produce the gamma-ray emission below the break energy is
\begin{linenomath}
\begin{equation}\label{eq:tdifflow}
\begin{aligned}
    t_{\mathrm{diff}}(E<E_C)&\sim \frac{(0.07\,\mathrm{pc})^2}{D_{\mathrm{LED}}(r=0.07\,\mathrm{pc})} \\
    &\approx 10^4\,(2.1\times 10^4)^{\frac{1-n}{3}}\,\hat{l}_c^{-\frac{2}{3}}\,\hat{B}^{\frac{1}{3}}\left(\frac{E}{\mathrm{TeV}}\right)^{-\frac{1}{3}} \,\mathrm{yr}.
\end{aligned}
\end{equation}
\end{linenomath}
On the other hand, for $E>E_C$, most of the emission occurs near the transition radius $r_C(E)$ from the LED to the HED regime, for which the diffusion time is found to be
\begin{linenomath}
\begin{equation}\label{eq:tdiffhigh}
\begin{aligned}
    t_{\mathrm{diff}}(E>E_C)&\sim\frac{(r_C(E))^2}{D_{\mathrm{LED}}(r_C(E))} \\ 
    &\approx 1.7\times 10^5\,(5.7)^{\frac{2-2n+m}{n-m}} (2.1\times 10^4)^{\frac{(1-n)(1-m)}{n-m}}\,\hat{l}_c^{\frac{2-n}{n-m}}\,\hat{B}^{\frac{2-m}{n-m}} \\ 
    &\times\left(\frac{E}{\mathrm{TeV}}\right)^{-\frac{2-m}{n-m}}\,\mathrm{yr}.
\end{aligned}
\end{equation}
\end{linenomath}
For any given gamma-ray energy $E_\gamma$, the emission currently observed roughly averages over the injection rate of CRs of energy $E\sim E_\gamma/\kappa_\pi$ over the last $t_{\mathrm{diff}}(E)$. As expected, for all relevant energies, these diffusion times are much shorter than those corresponding to the CMZ, therefore the gamma-ray emission of the point source is sensitive to much more recent activity (or inactivity) of the central black hole than the diffuse emission. Thus, the injection rate inferred from the point-source emission must not necessarily be the same as that explaining the CMZ, although we found them to be consistent with each other (with sizeable error bars on both).

It can also be seen from Eq. \ref{eq:tdiffhigh} that, for $E>E_C$, $t_{\mathrm{diff}}$ is a quickly decreasing function of $E$, mostly because of the decreasing critical radius $r_C(E)$ (Eq. \ref{eqn:crit radius}), which in turn is due to the rapidly increasing diffusion coefficient in the HED regime. For parameter choices favored by our model ($n=1$, $m=0$, $\hat l_c=3$, $\hat B=1$), we obtain $t_{\mathrm{diff}}\sim 0.5\,\mathrm{yr}(E/\mathrm{PeV})^{-2}$. Therefore, if the CR injection rate at PeV energies varies significantly on a comparable time-scale (i.e., of the order of a few months), as suggested by the enhanced activity of Sgr~A* in the radio, near-infrared, and X-ray bands during 2019 (\citealt{Weldon_2023} and references therein), this should be reflected in a significant variation of the gamma-ray emission at $E_\gamma\gtrsim 100\,\mathrm{TeV}$. This variable gamma-ray emission might be detectable in the near future by the Cherenkov Telescope Array Observatory (CTAO; \citealt{CTA_Consortium_2019,Viana_2019}).

\section{Conclusions}\label{sec:Conclusions}

The spatial distribution of the diffuse gamma-ray emission detected in the CMZ (corresponding to the inner $\sim$ 100 parsecs of the Milky Way), along with its extended spectrum reaching $\sim$ 100 TeV energies, have suggested the presence in the GC of a `PeVatron', that is, a CR accelerator capable of reaching PeV energies, which has been associated to the supermassive black hole Sgr~A* \citep{Abramowski_2016}. In addition, due to the apparently coincident position of the point-like gamma-ray source HESS~J1745$-$290 with Sgr~A*, it appears as an interesting possibility to explain both 
the diffuse CMZ emission and 
the point source as 
being due to the same CRs accelerated in the immediate vicinity of Sgr~A* and diffusing outwards from it. 
This scenario, however, is challenged by the fact that the spectrum of the point source shows a power-law behavior with a spectral turnover at a few TeV, not shown by the diffuse emission in the CMZ. Although this turnover is usually characterized as an exponential cutoff, the spectrum of the point source is also compatible with a broken power-law \citep{Aharonian_2009,Adams_2021}, as we also show in Appendix \ref{ap:fit_hess2016} using observations from \cite{Abramowski_2016}. 

In order to reconcile the CMZ and point source spectra, we propose a model for the point source in which the CRs are continuously injected near Sgr~A* (within a radius $r\sim 10 R_g$) and subsequently diffuse through its accretion flow ($r \lesssim 0.1$ pc). This way, very-high energy gamma rays are emitted in the accretion flow via inelastic hadronic collisions between the CR protons and the background protons. A key feature of this model is the existence of two CR diffusion regimes within the accretion flow of Sgr A*. These regimes, according to theoretical arguments and test particle simulations of CR propagation in synthetic MHD turbulence (which we assume is strong and has a Kolmogorov spectrum), depend on the ratio between the Larmor radius, $R_L$, of the CRs and the coherence length, $l_c$, of the turbulence. This way, the transition between these two regimes gives rise to a significant depletion of the highest energy CRs ($R_L/l_c \gg 1$) within the emission region, which explains the existence of a break in the point source spectrum. 

The free parameters of our model characterize the spectrum of the injected CRs, their propagation efficiency through the accretion flow, and the properties of the background gas.  Interestingly, the values of these parameters required to fit HESS J1745-290 are all consistent with expectations from the observations of the diffuse emission from the CMZ and with previous hydrodynamical and MHD simulations of the Sgr~A* accretion flow. The only exception is given by the (very uncertain) coherence length of the magnetic turbulence which needs to have an approximately homogeneous value, $l_c \sim (1-3)\times 10^{14}\,\mathrm{cm}\approx (3 \times 10^{-5} - 10^{-4})\,\mathrm{pc}$. Although disentangling the possible origin of this rather small coherence length is beyond the scope of this work, we speculate that its value might be affected by various processes, such as hydrodynamic instabilities in the colliding winds of the WR stars \citep{2020MNRAS.493..447C} or even MHD instabilities driven by the CRs themselves. Indeed, in \S \ref{sec:pressure} we found that the CR pressure can be dynamically important within the feeding region ($0.07\,\textrm{pc}\lesssim r \lesssim 0.4\,\textrm{pc}$), where gas injection from the WR stars occurs. This region thus appears as an ideal environment for the action of, e. g., the non-resonant instability, which is driven by the electric current of the CRs and has the potential to produce highly nonlinear MHD turbulence \citep{2004MNRAS.353..550B,2005MNRAS.358..181B,2009ApJ...694..626R,2010ApJ...717.1054R}. We defer the study of these possible sources of turbulence to future research.

Our results, therefore, support the hypothesis that Sgr~A* is capable of accelerating CRs up to a few PeV, contributing to explaining the origin of Galactic CRs up to the ``knee''. It is worth emphasizing, however, that although we have defined our CR acceleration region very close to Sgr A* (at $r \sim 10 R_g$), the obtained emission spectra from our model are highly insensitive to the precise location of the injection region, as long as this region is anywhere in the range $r\sim (10-10^3)\,R_g$ (see discussion in \S \ref{sec:qualitative}).

Our model has the potential to be tested by future TeV telescopes, such as CTAO \citep{CTA_Consortium_2019}. Given the small size of the emitting region ($r \lesssim 0.1$ pc, equivalent to $\lesssim 3$ arcsec in angular size), our model predicts a gamma-ray source that can not be resolved by any foreseeable TeV observatory. However, an interesting prediction of our model is that the spectrum of the point source should be a broken power-law, with specific spectral indices $\alpha_{LE}\approx 2.3$ for $E_{\gamma} \lesssim 6$ TeV and $\alpha_{HE}\approx 3$ for $E_{\gamma} \gtrsim 10$ TeV (see discussion in \S \ref{sec:simple}). This is in contrast to a single power-law with an exponential cutoff, as has been suggested by other models \citep[e.~g.,][]{Guo_2017}. This means that, at gamma-ray energies $\gtrsim$ 10 TeV, the gamma-ray spectrum obtained from our model becomes notably different from other competing models. This is particularly interesting given that CTAO will be $\sim 10$ times more sensitive than H.E.S.S., thus allowing, in principle, to discriminate between the different models. Further model discrimination might be done using the possible time variability of HESS~J1745$-$290 at the highest energies. This is interesting given the significant variability exhibited by Sgr~A* at various wavelengths. Although our model assumes a steady injection of CRs, in \S~\ref{sec:Discussion} we estimated that, if CR injection were variable, our model predicts potentially significant variability at $E_{\gamma} \sim 100$ TeV on time-scales as short as months. To date, however, observations have failed to detect variability of the HESS~J1745$-$290 gamma-ray point source, possibly because current observatories do not possess the required sensitivity to detect such variability. In this respect, the upcoming observations with CTAO may, in principle, have enough sensitivity to detect the point source variability. We defer the study of this possibility to future research.

\section*{Acknowledgements}

The authors gratefully acknowledge support from ANID-FONDECYT grants 1191673 and 1201582, as well as the Center for Astrophysics and Associated Technologies (CATA; ANID Basal grant FB210003). 

\bibliographystyle{aa}
\bibliography{paper}


\begin{appendix}
\section{Fitting HESS J1745-290 spectrum with a broken power-law}\label{ap:fit_hess2016}  

The central point source HESS J1745-290 is usually characterized by a hard power law spectrum with a photon index $\sim 2.1$ and an exponential cutoff at $\sim 10$ TeV. When compared with a pure power law, the power law with an exponential cutoff (ECPL) is clearly preferred by the data \citep{Abramowski_2016}.

Besides a power law with an exponential cutoff, the spectral shape of HESS J1745-290 has also been shown to be compatible with a broken power law with the break energy at a few TeV \citep{Aharonian_2009, Adams_2021}. Here we use more up-to-date HESS observations of this source obtained from 
\cite{Abramowski_2016} to confirm this point. 

\begin{figure}[h]
    \centering
    \includegraphics[width=0.46 \textwidth]{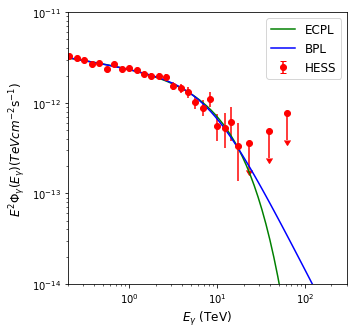}
      \caption{ \small  Least squares fits of the point source spectrum. The blue and green lines correspond to a broken power-law (BPL) and single power-law with exponential cutoff (ECPL) models for photon fluxes given by Eqs. \ref{eq:blp} and \ref{eq:ecpl}, respectively. The red circles correspond to HESS data from \cite{Abramowski_2016}.}
    \label{fig:BplVsEcpl}
\end{figure}

Fig. \ref{fig:BplVsEcpl} shows two least-squares fits of the point source (for which we used the Levenberg-Marquardt algorithm). The blue line corresponds to a broken power-law (BPL) model for the photon flux, $\Phi_{\gamma}(E_\gamma)$, given by
\begin{linenomath}
\begin{equation}
\Phi_{\gamma}(E_\gamma) = \frac{\Phi_0(E_\gamma/E_c)^{-\Gamma_1}}{1+(E_{\gamma}/E_c)^{\Gamma_2 -\Gamma_1}},
\label{eq:blp}
\end{equation}
\end{linenomath}
for which we obtain $\Phi_0 = (2.44\pm 1.2)\times 10^{-14}$ TeV$^{-1}$\,cm$^{-2}$\,s$^{-1}$, $E_c=8.22 \pm 1.4$ TeV, $\Gamma_1=2.18 \pm 0.04$, $\Gamma_2=3.89 \pm 0.52$, a p-value of $p=0.63$, and $\chi^2/$DOF$=0.87$. 

The green line corresponds to a ECPL model given by
\begin{linenomath}
\begin{equation}
\Phi_{\gamma}(E_\gamma) = \Phi_0(E_\gamma/E_c)^{-\Gamma_1}e^{-E_\gamma/E_c},
\label{eq:ecpl}
\end{equation}
\end{linenomath}
for which we find $\Phi_0 = (1.83\pm 0.88)\times 10^{-14}$ TeV$^{-1}$\,cm$^{-2}$\,s$^{-1}$, $E_c=10.13 \pm 1.89$ TeV, $\Gamma_1=2.13 \pm 0.03$, a p-value of $p=0.62$, and $\chi^2/$DOF$=0.88$. The similar values of $p$ and $\chi^2/$DOF obtained from these two fits imply that the BPL and ECPL models are both compatible with the current HESS data from \cite{Abramowski_2016}.

\section{Heuristic derivation of the diffusion coefficients}\label{ap:heuristic}  

Here, we give simple, rough, physical derivations of the effective mean free paths, $\lambda_{\mathrm{mfp}}$, and therefore of the diffusion coefficients, $D=\lambda_{\mathrm{mfp}}c/3$, of relativistic, charged particles in a medium with isotropic, magnetic Kolmogorov turbulence with coherence length $l_c$ and Larmor radius $R_L$. For the HED regime, $R_L\gg l_c$, a similar derivation has been given by \citet{Subedi_2017}, and we give it here for completeness. For the LED regime, $R_L\ll l_c$, we are not aware of a similar derivation in the literature. For simplicity and given the rough approximations involved, we generally ignore numerical factors $\sim 2\pi$ and smaller.

\subsection{High-energy diffusion (HED) regime}

If $R_L\gg l_c$, as a particle crosses a coherence length $l_c$, it deviates only by a small angle, $\theta_1\sim l_c/R_L$. These deviations are random, in different directions for successive steps of size $l_c$. Therefore, the direction of motion undergoes a random walk, accumulating a typical deviation $\theta_N\sim\sqrt{N}\theta_1$ after $N$ such steps. The direction of motion will have changed significantly once $\theta_N\sim 1,$ implying $N\sim(R_L/l_c)^2$, which defines the mean free path $\lambda_{\mathrm{mfp}}\sim N l_c\sim R_L^2/l_c$.

\subsection{Low-energy diffusion (LED) regime}

On scales $\ll l_c$, the magnetic field can be approximated by $\vec B=\vec B_0+\delta\vec B(\vec r)$, where $\vec B_0=\mathrm{constant}$ and $|\delta\vec B(\vec r)|\ll B_0$. If $R_L\equiv\gamma mc^2/(qB_0)\ll l_c$, where $\gamma$, $m$, and $q$ are the Lorentz factor, the mass, and the charge of the particles, respectively, the latter will follow a roughly helical motion, typically with velocity components of similar magnitude (not much smaller than $c$) along $\vec B_0$, $\vec v_\parallel$, and perpendicular to it, $\vec v_\perp$. Thus, the particles will move a distance $\sim R_L$ along $\vec B_0$ while completing an orbit of radius $\sim R_L$ in the perpendicular plane. 

The equation of motion for the parallel velocity component is
\begin{linenomath}
\begin{equation}\label{eq:v_par}
    \gamma m\frac{\mathrm{d}\vec v_\parallel}{\mathrm{d}t}=\frac{q}{c}\vec v_\perp\times\delta\vec B.
\end{equation}
\end{linenomath}
Since the direction of $\vec v_\perp$ completes a loop while the particle travels a distance $\sim R_L$, $\vec v_\parallel$ will be mostly affected by field perturbations on roughly the same scale, which we will call $\delta B^*$. Its typical change while traveling this scale will be small, $\Delta v_{\parallel,1}\sim c\delta B^*/B_0$, but these random changes can accumulate, yielding a substantial change $\Delta v_{\parallel,N}\sim\sqrt{N}\Delta v_{\parallel,1}\sim c$ after $N\sim(c/\Delta v_{\parallel,1})^2$ steps, corresponding to a one-dimensional mean free path (along $\vec B_0$) given by $\lambda_{\mathrm{mfp}}\sim R_L\left(B_0/\delta B^*\right)^2$. For Kolmogorov turbulence, $B_0/\delta B^*\sim(l_c/R_L)^{1/3}$, therefore the one-dimensional mean free path is 
\begin{linenomath}
\begin{equation}
    \lambda_{\mathrm{mfp}}\sim R_L^{1/3}l_c^{2/3}, 
\end{equation}
\end{linenomath}
with a corresponding diffusion coefficient $D_{\mathrm{1D}}=\lambda_{\mathrm{mfp}}c/3$. 

The time required to diffuse a distance $l_c$ along $\vec B_0$ is $t_c\sim l_c^2/D_{\mathrm{1D}}$. On each patch of size $l_c$, the magnetic field has a random orientation. Thus, on larger scales, the diffusion process becomes three-dimensional, with steps of size $l_c$ traversed at a typical speed $v_c\sim l_c/t_c$, so the large-scale, three-dimensional diffusion coefficient is $D\sim l_cv_c/3\sim l_c^2/t_c\sim D_{\mathrm{1D}},$ as in the one-dimensional analysis above.

\section{Full expressions for $\mathrm{d}n_{\mathrm{CR}}/\mathrm{d}E$}
\label{ap:fullexpressions}

In this appendix we describe the analytical treatment for the CR density distribution per unit energy $\mathrm{d}n_{\mathrm{CR}}/\mathrm{d}E$ of our full model described in \S\ref{sec:crdensity}.

\subsection{Continuity and boundary conditions}

\begin{figure}[h]
    \label{fig: cr dens model}
     \centering
	\includegraphics[scale=0.3]{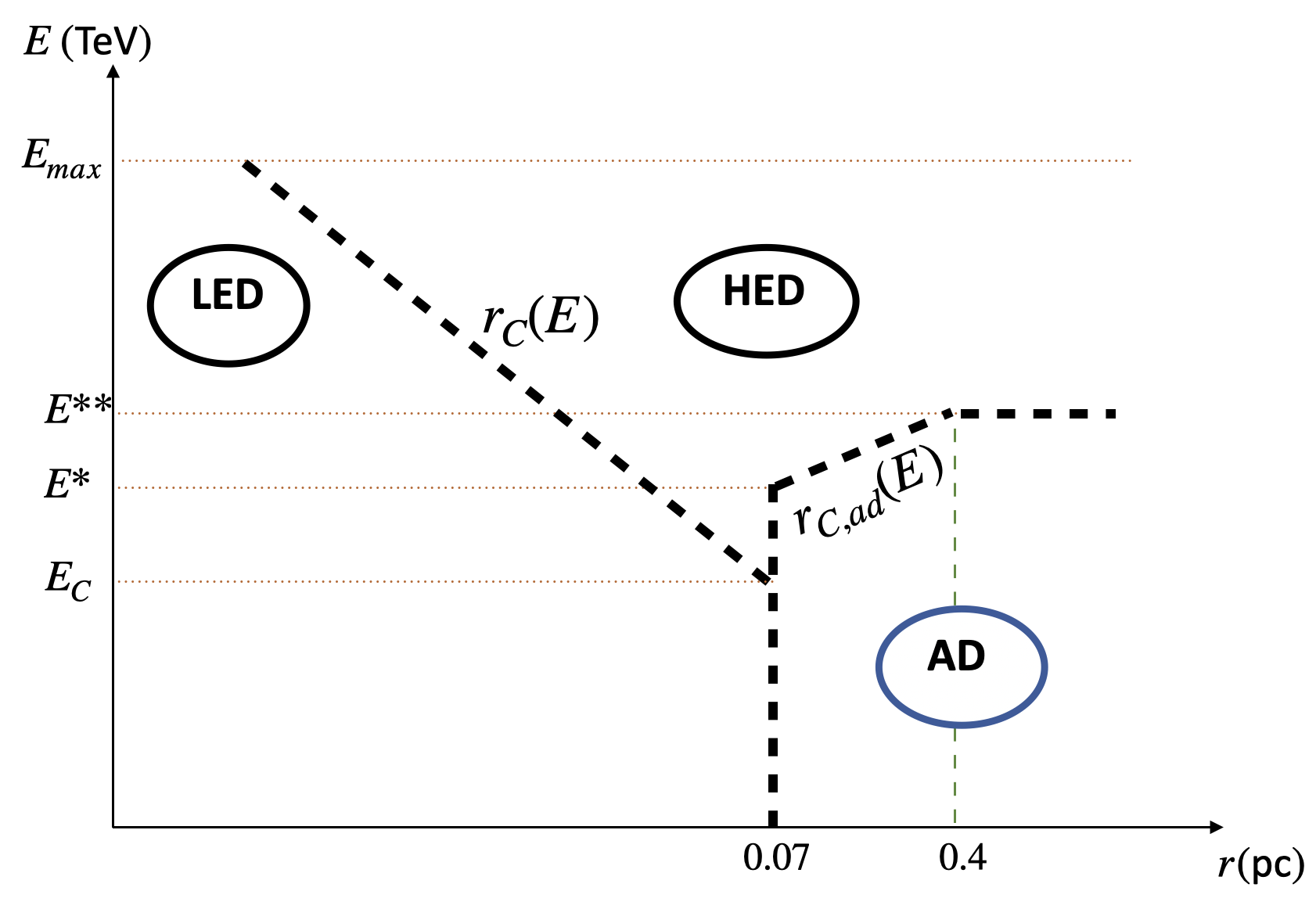}
    \caption{Schematic representation of the different CR transport regimes used to calculate $\mathrm{d}n_{\mathrm{CR}}/\mathrm{d}E$, as a function of radius $r$ and CR energy $E$. These regimes correspond to low-energy diffusion (LED), high-energy diffusion (HED) and advection (AD).}
	\end{figure}

As explained in \ref{sec:crinject}, in order to determine the CR density distribution on the Sgr A* accretion flow, we solve Eq. \ref{eqn1: diff} neglecting one of the two terms on the right hand side, depending on whether diffusion of advection dominates. In the case dominated by diffusion, we have
\begin{linenomath}
\begin{equation}
\label{eqn1: diffp}
\frac{\mathrm{d}N}{\mathrm{d}E\mathrm{d}t}(E)=-4\pi r^2 D(E,r) \frac{\partial}{\partial r} \left(\frac{\mathrm{d}n_{\mathrm{CR}}}{\mathrm{d}E}(E,r)\right),
\end{equation} 
\end{linenomath}
\noindent while is advection is the dominant transport mechanism of CRs, 
\begin{linenomath}
\begin{equation}
\label{eqn2: adv}
\frac{\mathrm{d}N}{\mathrm{d}E\mathrm{d}t}(E)=4\pi r^2 \: v_{\textrm{gas}}(r)\: \frac{\mathrm{d}n_{\textrm{CR}}}{\mathrm{d}E}(E,r).
\end{equation}
\end{linenomath}
The above equations for $\mathrm{d}n_{\mathrm{CR}}/\mathrm{d}E$ are constrained by continuity restrictions, which we characterize in Fig. \ref{fig: cr dens model}, noting that: $r_C(E)$ represent the critical radius at which a CR of energy $E$ will transit from the LED regime to HED regime (see Eq. \ref{eqn:crit radius}), $r_{C,ad}(E)$ represent the radius at which a CR of energy $E$ transit from the HED regime to the AD regime (see Eq. \ref{eq.v_gas=v_diff}), given that the gas velocity is determined by Eq. \ref{eq. v_ad}. 

Fig \ref{fig: cr dens model}, shows a schematic representation of the different CR transport regimes used to calculate $\mathrm{d}n_{\mathrm{CR}}/\mathrm{d}E$, as a function of radius $r$ and CR energy $E$. We see that, besides $E_C$ (defined by $r_C(E)=0.07$ pc), the limits between these regions are characterized by the energies $E^*$ and $E^{**}$, corresponding to the energies where $r_{C,ad}(E)$ becomes equal to 0.07 pc and 0.4 pc, respectively. Thus, the continuity conditions for the CR density can be described within the four energy ranges shown in Fig \ref{fig: cr dens model}:
\begin{enumerate}

    \item Energy range $E\,\epsilon \,[1\text{TeV}-E_C]$: 
    \begin{equation}
    \frac{dn_{CR}}{dE}|_{LED}(E,r=0.07\text{pc})=\frac{dn_{CR}}{dE}|_{AD}(E,r= 0.07\text{pc}).
    \end{equation}
    
    \item Energy range $E\,\epsilon \,[E_C-E^*]$:
    \begin{equation}
    \frac{dn_{CR}}{dE}|_{LED}(E,r=r_C(E))=\frac{dn_{CR}}{dE}|_{HED}(E,r=r_C(E)),\text{ and}
    \end{equation} 
    \begin{equation}
      \frac{dn_{CR}}{dE}|_{HED}(E,r=0.07\text{pc})=\frac{dn_{CR}}{dE}|_{AD}(E,r=0.07\text{pc}).  
    \end{equation}
    \item Energy range $E\,\epsilon \,[E^*-E^{**}]$: 
    \begin{equation}
    \frac{dn_{CR}}{dE}|_{LED}(E,r=r_C(E))=\frac{dn_{CR}}{dE}|_{HED}(E,r=r_C(E)),\text{and}
    \end{equation}  
    \begin{equation}
    \frac{dn_{CR}}{dE}|_{HED}(E,r=r_{C,ad}(E))=\frac{dn_{CR}}{dE}|_{AD}(E,r=r_{C,ad}(E))
    \end{equation}
    
    \item Energy range $E\,\epsilon \,[E^{**}-E_{max}]$: 
   \begin{equation}
    \frac{dn_{CR}}{dE}|_{LED}(E,r=r_C(E))=\frac{dn_{CR}}{dE}|_{HED}(E,r=r_C(E)).
    \end{equation}
\end{enumerate}
\medskip
In all cases, we impose $dn_{CR}/dE (E,r \rightarrow \infty)\rightarrow 0$. 

\subsection{Full model for the CR density distribution}
Here we show the full expression for $dn_{CR}/dE$ obtained from solving Eqs. \ref{eqn1: diffp} and \ref{eqn2: adv} under the aforementioned restrictions, for the four different energy ranges depicted in Fig. \ref{fig: cr dens model} and for the different radial ranges within them.

\begin{enumerate}
\item Energy range $E\,\epsilon \,[1\text{TeV}-E_C]$: \\

    \medskip
    
In the case of $r\leq0.07$ pc,
\begin{linenomath}
    \begin{equation}
       \begin{aligned}
     \frac{\mathrm{d}n_{\mathrm{CR}}}{\mathrm{d}E}(E,r)= & \,3 \: (2.1\times 10^4)^{2m/3}\,\hat{Q} \,  \hat{l}_c^{-2/3}\,\hat{B}^{\frac{1}{3}} \,\frac{f(q,E_{\textrm{max}})}{1+\frac{1}{3}(n+2m)}\\[1.5ex]
     &   \times \, \left(\frac{E}{1 \text{TeV}}\right)^{-q-\frac{1}{3}} \left(\frac{r}{10 R_g}\right)^{-(1+\frac{1}{3}(n+2m))} \:\text{erg}^{-1}\text{cm}^{-3}\\[1.5ex]
     & +3\times 10^{-7}\hat{Q}f(q,E_{\textrm{max}})\left(\frac{E}{1 \text{TeV}}\right)^{-q}\,\text{erg}^{-1}\text{cm}^{-3}\\[1.5ex]
     & -3 \: (2.1\times 10^4)^{-(1+n/3)}\,\hat{Q} \,  \hat{l}_c^{-2/3}\,\hat{B}^{\frac{1}{3}} \,\frac{f(q,E_{\textrm{max}})}{1+\frac{1}{3}(n+2m)}\\[1.5ex]
     &\times \, \left(\frac{E}{1 \text{TeV}}\right)^{-q-\frac{1}{3}} \:\text{erg}^{-1}\text{cm}^{-3},\\
     \end{aligned}  
    \end{equation}
\end{linenomath}
where the first term on the right hand is the same as the right hand side of Eq. \ref{Eq: CR dens1}, in which boundary condition terms were neglected. In the cases $0.07$ pc $\leq r\leq 0.4$ pc and $r\geq 0.4$ pc, the expressions for $\mathrm{d}n_{\mathrm{CR}}/\mathrm{d}E$ are the same as Eqs. \ref{eqn:Adv1} and \ref{eqn:Adv2}, respectively, since no boundary terms were neclected in those cases.
\medskip
\item Energy range $E\,\epsilon \,[E_C-E^*]$:

If $r\leq r_{C}(E)$,
\begin{linenomath}
\begin{equation}
    \begin{aligned}
     \frac{\mathrm{d}n_{\mathrm{CR}}}{\mathrm{d}E}(E,r)= & \,3 \: (2.1\times 10^4)^{2m/3}\,\hat{Q} \,  \hat{l}_c^{-2/3}\,\hat{B}^{\frac{1}{3}} \,\frac{f(q,E_{\textrm{max}})}{1+\frac{1}{3}(n+2m)}\\
     &   \times \, \left(\frac{E}{1 \text{TeV}}\right)^{-q-\frac{1}{3}} \left(\frac{r}{10 R_g}\right)^{-(1+\frac{1}{3}(n+2m))} \:\text{erg}^{-1}\text{cm}^{-3}\\[1.5ex]
     &+10^{9}  \,(2.1\times 10^4)^{\frac{m(1+n)}{n-m}}\: \hat{Q} \: \hat{l}_c^{\frac{-(1+n)}{n-m}} \:  \hat{B}^{\frac{-(1+m)}{n-m}} \\[1.5ex]
     & \times \frac{f(q,E_{\textrm{max}})}{1+2n-m}(1.3\times 10^5)^\frac{-(1+2n-m)}{n-m}\\
     &\times \left(\frac{E}{1 \text{TeV}}\right)^{\frac{m+1}{n-m}-q}  \:\text{erg}^{-1}\text{cm}^{-3}\\[1.5ex]
     &-\,3 \: \,(2.1\times 10^4)^{\frac{m(1+n)}{n-m}}\: \hat{Q} \: \hat{l}_c^{\frac{-(1+n)}{n-m}} \:  \hat{B}^{\frac{-(1+m)}{n-m}}\\
     &\times \frac{f(q,E_{\textrm{max}})}{1+\frac{1}{3}(n+2m)}(1.3\times 10^5)^\frac{-(1+\frac{1}{3}(n+2m))}{n-m}\\[1.5ex]
     &\times \, \left(\frac{E}{1 \text{TeV}}\right)^{\frac{m+1}{n-m}-q}  \:\text{erg}^{-1}\text{cm}^{-3}\\[1.5ex]
     & +3\times 10^{-7}\hat{Q}f(q,E_{\textrm{max}})\left(\frac{E}{1 \text{TeV}}\right)^{-q}\,\text{erg}^{-1}\text{cm}^{-3}\\[1.5ex]
     & -10^{9}  \,(2.1\times 10^4)^{-(1+2n)}\: \hat{Q} \: \hat{l}_c \:  \hat{B}^{2}\\ &\times \frac{f(q,E_{\textrm{max}})}{1+2n-m} \, \left(\frac{E}{1 \text{TeV}}\right)^{-q-2}  \text{erg}^{-1}\text{cm}^{-3},\\[1.5ex]
     \end{aligned}  
\end{equation}
\end{linenomath}
where, the first term on the right hand side is again equal to the right hand side of the simpler Eq. \ref{Eq: CR dens1}. \newline

\noindent If $r_{C}(E)\leq r\leq 0.07$ pc,
\begin{linenomath}
\begin{equation}
    \begin{aligned}
    \frac{\mathrm{d}n_{\mathrm{CR}}}{\mathrm{d}E}(E,r)= & \, 10^{9}  \,(2.1\times 10^4)^{-m}\: \hat{Q} \: \hat{l}_c \:  \hat{B}^{2}\, \frac{f(q,E_{\textrm{max}})}{1+2n-m} \\[1.5ex]
     & \times \, \left(\frac{E}{1 \text{TeV}}\right)^{-q-2}\left(\frac{r}{10 R_g}\right)^{-(1+2n-m)}\: \text{erg}^{-1}\text{cm}^{-3}\\[1.5ex]
    & +3\times 10^{-7}\hat{Q}f(q,E_{\textrm{max}})\left(\frac{E}{1 \text{TeV}}\right)^{-q}\,\text{erg}^{-1}\text{cm}^{-3}\\[1.5ex]
     & -10^{9}  \,(2.1\times 10^4)^{-(1+2n)}\: \hat{Q} \: \hat{l}_c \:  \hat{B}^{2}\, \frac{f(q,E_{\textrm{max}})}{1+2n-m} \\[1.5ex]
     & \times \, \left(\frac{E}{1 \text{TeV}}\right)^{-q-2}  \text{erg}^{-1}\text{cm}^{-3},\\[1.5ex]
     \end{aligned}   
\end{equation}
\end{linenomath}
where the first term on the right hand side is the same as the right hand side of Eq. \ref{Eq: CR dens2}, which does not include boundary terms. In the cases $0.07$ pc $\leq r\leq 0.4$ pc and $r\geq 0.4$ pc, the expressions for $\mathrm{d}n_{\mathrm{CR}}/\mathrm{d}E$ are the same as those provided by Eqs. \ref{eqn:Adv1} and \ref{eqn:Adv2}, since in those cases no boundary terms were neglected. \newline
\medskip
\medskip

\item Energy range $E\,\epsilon \,[E^*-E^{**}]$:\newline

\noindent If $r\leq r_{C}(E)$

\begin{linenomath}
    \begin{equation}
        \begin{aligned}
        \frac{\mathrm{d}n_{\mathrm{CR}}}{\mathrm{d}E}(E,r)= & \,3 \: (2.1\times 10^4)^{2m/3}\,  \hat{l}_c^{-2/3}\,\hat{B}^{\frac{1}{3}} \,\frac{\hat{Q} \,f(q,E_{\textrm{max}})}{1+\frac{1}{3}(n+2m)}\\[1.5ex]
     &   \times \,          \left(\frac{E}{1 \text{TeV}}\right)^{-q-\frac{1}{3}} \left(\frac{r}{10 R_g}\right)^{-(1+\frac{1}{3}(n+2m))} \:\text{erg}^{-1}\text{cm}^{-3}\\[1.5ex]
     &+10^{9}  \,(2.1\times 10^4)^{\frac{-m(1+n)}{n-m}}\: \hat{l}_c^{\frac{-(1+n)}{n-m}} \:  \hat{B}^{\frac{-(1+m)}{n-m}}\,  \\[1.5ex]
     & \times \frac{\hat{Q} \:f(q,E_{\textrm{max}})}{1+2n-m}(1.3\times 10^5)^\frac{-(1+2n-m)}{n-m}\, \\
     &\times \left(\frac{E}{1 \text{TeV}}\right)^{\frac{m+1}{n-m}-q}  \:\text{erg}^{-1}\text{cm}^{-3}\\[1.5ex]
     &-\,3 \: \,(2.1\times 10^4)^{\frac{-m(1+n)}{n-m}}\: \hat{l}_c^{\frac{-(1+n)}{n-m}} \:  \hat{B}^{\frac{-(1+m)}{n-m}}\,(1.3\times 10^5)^\frac{-(1+\frac{1}{3}(n+2m))}{n-m} \\[1.5ex]
     &\times \frac{\hat{Q} \:f(q,E_{\textrm{max}})}{1+\frac{1}{3}(n+2m)}\, \left(\frac{E}{1 \text{TeV}}\right)^{\frac{m+1}{n-m}-q}  \:\text{erg}^{-1}\text{cm}^{-3}\\[1.5ex]
     & +3\times 10^6 \,(2.1\times10^4)^\frac{3m}{2n-m-2}\hat{l}_c^\frac{-3}{2n-m-2}\hat{B}^\frac{-6}{2n-m-2}
     \\[1.5ex]
     &\times \hat{Q}f(q,E_{\textrm{max}})(400)^{\frac{-3}{2n-m-2}}\,\left(\frac{E}{1 \text{TeV}}\right)^{\frac{6}{2n-m-2}-q}\,\text{erg}^{-1}\text{cm}^{-3}\\[1.5ex]
     & -10^{9} \,(2.1\times10^4)^\frac{3m}{2n-m-2}\hat{l}_c^\frac{-3}{2n-m-2}\hat{B}^\frac{-6}{2n-m-2} \\[1.5ex]
     &\times \frac{\hat{Q}\,f(q,E_{\textrm{max}})}{1+2n-m}(400)^{\frac{m-2n-1}{2n-m-2}}\, \left(\frac{E}{1 \text{TeV}}\right)^{\frac{6}{2n-m-2}-q} \text{erg}^{-1}\text{cm}^{-3},\\[1.5ex]
     \end{aligned}  
\end{equation}
\end{linenomath}
where the first term on the right hand side coincides with the right hand side of Eq. \ref{Eq: CR dens1}.\newline

\noindent If $r_{C}(E)\leq r \leq r_{C,ad}(E)$,
\begin{linenomath}
\begin{equation}
    \begin{aligned}
     \frac{\mathrm{d}n_{\mathrm{CR}}}{\mathrm{d}E}(E,r)= & \, 10^{9}  \,(2.1\times 10^4)^{-m}\: \hat{Q} \: \hat{l}_c \:  \hat{B}^{2}\, \frac{f(q,E_{\textrm{max}})}{1+2n-m} \\[1.5ex]
     & \times \, \left(\frac{E}{1 \text{TeV}}\right)^{-q-2}\left(\frac{r}{10 R_g}\right)^{-(1+2n-m)}\: \text{erg}^{-1}\text{cm}^{-3}\\[1.5ex]
    & +3\times 10^6\,(2.1\times10^4)^\frac{3m}{2n-m-2}\hat{l}_c^\frac{-3}{2n-m-2}\hat{B}^\frac{-6}{2n-m-2}
     \\[1.5ex]
     &\times \hat{Q}f(q,E_{\textrm{max}})(400)^{\frac{-3}{2n-m-2}}\,\left(\frac{E}{1 \text{TeV}}\right)^{\frac{6}{2n-m-2}-q}\,\text{erg}^{-1}\text{cm}^{-3}\\[1.5ex]
     & -10^{9} \,(2.1\times10^4)^\frac{3m}{2n-m-2}\hat{l}_c^\frac{-3}{2n-m-2}\hat{B}^\frac{-6}{2n-m-2} \\[1.5ex]
     &\times \frac{\hat{Q}\,f(q,E_{\textrm{max}})}{1+2n-m}(400)^{\frac{m-2n-1}{2n-m-2}}\,\left(\frac{E}{1 \text{TeV}}\right)^{\frac{6}{2n-m-2}-q} \text{erg}^{-1}\text{cm}^{-3},\\[1.5ex]
     \end{aligned}    
\end{equation}
\end{linenomath}
where the first term on the right hand side coincides with the right hand side of Eq. \ref{Eq: CR dens2}.\newline

\noindent If $r_{C,ad}(E)\leq r\leq 0.4$ pc and $r\geq 0.4$ pc, once again, the expressions for $\mathrm{d}n_{\mathrm{CR}}/\mathrm{d}E$ are the same as those provided in the main text for the ADV regime (Eqs. \ref{eqn:Adv1} and \ref{eqn:Adv2}, respectively).\newline

\item Finally, the energy range $E\,\epsilon\, [E^{**},E_{max}]$:\newline

\noindent If $r\leq r_{C}(E)$,
\begin{linenomath}
\begin{equation}
    \begin{aligned}
     \frac{\mathrm{d}n_{\mathrm{CR}}}{\mathrm{d}E}(E,r)= & \,3 \: (2.1\times 10^4)^{2m/3}\,\hat{Q} \,  \hat{l}_c^{-2/3}\,\hat{B}^{\frac{1}{3}} \,\frac{f(q,E_{\textrm{max}})}{1+\frac{1}{3}(n+2m)}\\[1.5ex]
     &   \times \, \left(\frac{E}{1 \text{TeV}}\right)^{-q-\frac{1}{3}} \left(\frac{r}{10 R_g}\right)^{-(1+\frac{1}{3}(n+2m))} \:\text{erg}^{-1}\text{cm}^{-3}\\[1.5ex]
     &+10^{9}  \,(2.1\times 10^4)^{\frac{-m(1+n)}{n-m}}\: \hat{l}_c^{\frac{-(1+n)}{n-m}} \:  \hat{B}^{\frac{-(1+m)}{n-m}}\, \\[1.5ex]
     & \times \frac{\hat{Q} \:f(q,E_{\textrm{max}})}{1+2n-m}(1.3\times 10^5)^\frac{-(1+2n-m)}{n-m}\\
     &\times \left(\frac{E}{1 \text{TeV}}\right)^{\frac{m+1}{n-m}-q}  \:\text{erg}^{-1}\text{cm}^{-3}\\[1.5ex]
     &-\,3 \: \,(2.1\times 10^4)^{\frac{-m(1+n)}{n-m}}\: \hat{l}_c^{\frac{-(1+n)}{n-m}} \:  \hat{B}^{\frac{-(1+m)}{n-m}}\\ &\times \frac{f(q,E_{\textrm{max}})}{1+\frac{1}{3}(n+2m)}(1.3\times 10^5)^\frac{-(1+\frac{1}{3}(n+2m))}{n-m}\\
     &\times \, \left(\frac{E}{1 \text{TeV}}\right)^{\frac{m+1}{n-m}-q}  \:\text{erg}^{-1}\text{cm}^{-3},\\[1.5ex]
     \end{aligned}  
\end{equation}
\end{linenomath}
where the first term on the right hand side is the same as the right hand side of Eq. \ref{Eq: CR dens1}. In the radial range $r\geq r_C(E)$, the expression for $\mathrm{d}n_{\mathrm{CR}}/\mathrm{d}E$ is the same as Eq. \ref{Eq: CR dens2}, since in the energy range $E\,\epsilon\, [E^{**},E_{max}]$, the CR density in the HED regime simply needs to satisfy the boundary condition $dn_{CR}/dE (E,r \rightarrow \infty)\rightarrow 0$, which is already satisfied by Eq. \ref{Eq: CR dens2}.
\end{enumerate}
\end{appendix}
%
%
\end{document}